\documentclass[preprint]{elsarticle}

\usepackage[utf8]{inputenc}
\usepackage[svgnames]{xcolor}
\definecolor{navyblue}{rgb}{0.0, 0.0, 0.5}
\usepackage{graphicx}
\usepackage{dcolumn}
\usepackage{bm}
\usepackage{float}
\usepackage{tikz}
\usepackage{pgfplots} 
\usetikzlibrary{spy}
\usepackage{caption}
\usepackage[margin=2.5cm]{geometry}
\pgfplotsset{compat=1.15}
\usepackage[mathscr]{euscript}
\usepackage{svg}
\usepackage{natbib}
\setcitestyle{super}
\usepackage{pgfplotstable}
\usepackage{gensymb}
\usepackage{newunicodechar}
\usepackage{parskip}
\usepackage{nomencl}
\usepackage{color, soul}
\usepackage{amsmath}
\usepackage{amssymb}
\usepackage{lineno,hyperref}
\modulolinenumbers[5]
\usepackage{textcomp}

\usepackage{siunitx}
\usepackage{booktabs}
\usepackage{multirow}
\usepackage{array}
\usepackage{tabularx}
\newcolumntype{L}{>{\centering\arraybackslash}m{6cm}}
\newcolumntype{J}{>{\centering\arraybackslash}m{6cm}}
\newcolumntype{K}{>{\centering\arraybackslash}m{1.5cm}}
\newcolumntype{Q}{>{\raggedleft\arraybackslash}X}
\newcolumntype{C}{>{\centering\arraybackslash}X}
\newcommand{\rowgroup}[1]{\hspace{-1em}#1}

\newcommand{\tagarray}{%
\mbox{}\refstepcounter{equation}%
$(\theequation)$%
}
\DeclareCaptionLabelFormat{AppendixTables}{A.#2}

\journal{Journal of Power Sources}

\begin{document}

\begin{frontmatter}

\title{A Computationally Informed Realisation Algorithm for Lithium-Ion Batteries Implemented with LiiBRA.jl}

\author[mymainaddress]{Brady Planden\corref{mycorrespondingauthor}}
\author[mymainaddress]{Katie Lukow}
\author[mymainaddress]{Paul Henshall}
\author[mymainaddress]{Gordana Collier}
\author[mymainaddress]{Denise Morrey}

\cortext[mycorrespondingauthor]{Corresponding author. Tel: +44 1865 647297. \\ Email address: brady.planden@brookes.ac.uk}

\address[mymainaddress]{School of Engineering, Computing, and Mathematics, Oxford Brookes University, Wheatley Campus, Wheatley, Oxford, OX33 1HX }

\begin{abstract}
{Real-time battery modelling advancements have quickly become a requirement as the adoption of battery electric vehicles (BEVs) has rapidly increased. In this paper an open-source, improved discrete realisation algorithm, implemented in Julia for creation and simulation of reduced-order, real-time capable physics-based models is presented. This work reduces the Doyle-Fuller-Newman electrochemical model into continuous-form transfer functions and introduces a computationally informed discrete realisation algorithm (CI-DRA) to generate the reduced-order models. Further improvements in conventional offline model creation are obtained as well as achieving in-vehicle capable model creation for ARM based computing architectures. Furthermore, a sensitivity analysis on the resultant computational time is completed as well as experimental validation of a worldwide harmonised light vehicle test procedure (WLTP) for a LG Chem. M50 21700 parameterisation. A performance comparison to the conventional Matlab implemented discrete realisation algorithm (DRA) is completed showcasing a mean computational time improvement of 88\%. Finally, an ARM based compilation is investigated for in-vehicle model generation and shows a modest performance reduction of 43\% when compared to the x86 implementation while still generating accurate models within 5.5 seconds. }
\end{abstract}


\end{frontmatter}


\section{\label{sec:Intro}Introduction}
{
With the rapidly increasing adoption of battery electric vehicles (BEVs), improvements with in-vehicle battery modelling and control are required to improve safety and driving performance, while ensuring the vehicle battery pack reaches the desired lifetime. Providing a viable method for capturing real-time degradation mechanisms coupled with physics-based electrochemical models is a key achievement required for future electric vehicle advancements.\cite{edge_lithium_2021} In-vehicle battery control systems, better know as battery management systems (BMS), ensure hardware limits are maintained while providing the requested interaction from the operator. These systems accomplish this by ensuring the pack is in a safe state for operation, protecting the individual cells from abuse and reducing the battery pack degradation over the lifetime of operation. The BMS relies heavily on online predictive models that are utilised for hardware limit forecasting, plant-based control structures, and state estimation to achieve this. Data-driven models such as equivalent circuit models (ECM)\cite{zhao_observability_2017,hu_comparative_2012,plett_battery_2015} are commonly utilised for this prediction as they provide reasonable performance and have a well established path for model creation. These models are numerically deployed onto in-vehicle embedded systems and provide information to the BMS that typically would not be attainable via direct sensing methods. This information is provided at designated non-flexible time intervals to the onboard control strategy with key performance indicators calculated such as state-of-power (SOP), state-of-charge (SOC), and state-of-health (SOH). Each of these state variables provide insight on the vehicle's capabilities for future operation.\\

These data-driven battery models can provide a fast, reliable solution; however, their creation requires existing data that encompasses the entire operating range of the cell to ensure a stable response for the predicted operating conditions. Obtaining this data is not only time consuming, on the order of multiple months of test channel time, but also requires expensive test equipment. These models also tend to lack electrochemical generality due to their nature and the model data requirements needed to achieve acceptable performance. As an example, ECMs utilise idealised, theoretical electrical components to represent cell behaviour, whose properties are numerically calibrated such that the model output is consistent with only a few basic measured cell characteristics, such as terminal voltage.  As such, generality isn't achievable for cell characteristics across varying chemistries, geometries, and operating conditions. Additionally, without observing internal electrochemical states during data acquisition, insight into these properties of the cell are not available, making prediction of long-term battery pack degradation difficult and inaccurate. \cite{edge_lithium_2021,petit_development_2016,de_hoog_combined_2017,reniers_improving_2018,schimpe_comprehensive_2018} \\

An alternative to data-driven models are physics based models, such as the Doyle-Fuller-Newman Pseudo-2D (DFN) \cite{doyle_modeling_1993, fuller_simulation_1994} or the Single Particle Model (SPM)\cite{marquis_asymptotic_2019}. These models provide internal electrochemical insight and can offer a viable solution for degradation-sensitive next generation cells such as anode-free liquid electrolyte lithium-metal.\cite{jang_towards_2021, mishra_perspectivemass_2021, reniers_review_2019} Furthermore, accurate long-term predictions are within these models capabilities, providing coupling for cell degradation mechanisms such as intercalation electrode lithium plating, loss of active material (LAM) and lithium inventory (LLI), pore clogging, and dendrite growth.\cite{edge_lithium_2021, reniers_review_2019} However, this coupling can be mathematically complex, and requires knowledge of multiple physical parameters which can be difficult and/or expensive to obtain. Due to the beneficial information provided by physics-based models, work has been completed to reduce the numerical complexity and computational performance requirements. Simplification of the partial differential equations governing the system is one such method and has resulted in the SPM and its electrolyte capable (SPMe) form.\cite{marquis_asymptotic_2019} Additional methods include, Pad\'e approximations, \cite{Forman_2011} residue grouping,\cite{smith_model_2008,Ramadesigan_2010} and parabolic solid-phase diffusion approximations.\cite{Subramanian_2005} To achieve deployment on battery management systems, further reduction is required. One such reduction method reduces the partial differential equations to continuous-form transfer functions combined with eigensystem realisation algorithms.\cite{lee_discrete-time_2012,plett_battery_2015,rodriguez_comparing_2019} Likewise, Jin et al developed a reduced-order capacity-loss model for graphite anodes, that focused on only the most significant degradation mechanisms to improve computational efficiency.\cite{jin_physically-based_2017} Similarly, Han et al developed a reduced order lumped electrochemical-thermal cell model by applying a state space approach to transform partial differential equations into ordinary differential equations.\cite{2021JPS...49029571H} These reductions aim to deploy capable predictive models to battery management systems, and are heavily numerically reduced. \\

As discussed above, there are numerous benefits to deploying these reduced-order electrochemical models onto a BMS, including improved accuracy for predictions in SOC, SOP, and SOH;\cite{SMITH2006662,Comp_Framework_MPC} however, to ensure robust, stable operation of the BMS, the deployed model needs to be real-time capable for the given hardware. This requirement is fulfilled if the online model can be solved before the BMS is required to communicate the solution, or provide a control interaction. Depending on the application, this solution rate can have requirements as low as 1Hz to upwards of 10Hz in fast dynamic systems. Therefore, the final reduced model must be capable within these ranges to be seen as a viable solution. In this paper, a novel software package developed in Julia\cite{Julia-2017} is presented for generating numerically reduced real-time capable physics-based models. This work presents an improved realisation algorithm for fast solution generation as well as investigates in-vehicle model creation as a viable method for degradation informed models. Finally, a sensitivity analysis is performed alongside numerical verification and experimental validation. \\
}

\newpage

\begin{table}[!th]
\centering
\renewcommand{\arraystretch}{1.3}
\begin{tabularx}{0.7\textwidth}{| >{\hsize=0.2\hsize\linewidth=\hsize}C  
>{\hsize=0.8\hsize\linewidth=\hsize}C |}
\hline
\textbf{Nomenclature} & \\
$n$ & Negative domain \\
$s$ & Separator \\
$p$ & Positive domain \\
$L_k$ & Domain length, $k \in \{n,s,p\}$  \\
$^*$ & Dimensionless operator \\
$\phi_{s,k}$ & Solid potential, $k \in \{n,s,p\}$ \\
$\phi_{e,k}$ & Electrolyte potential, $k \in \{n,s,p\}$ \\
$c_{s,k}$ & Solid lithium concentration, $k \in \{n,s,p\}$ \\
$c_{e,k}$ & Electrolyte lithium concentration, $k \in \{n,s,p\}$ \\
$i_{e,k}$ & Ionic current density, $k \in \{n,s,p\}$ \\
$N_{e,k}$ & Electrolyte molar flux, $k \in \{n,s,p\}$ \\
$\sigma_{k}$ & Solid conductivity, $k \in \{n,s,p\}$ \\
$\kappa_{e}$ & Electrolyte conductivity \\
$I_{app}$ & Applied current density \\
$R_s$ & Particle radius \\
$D_s$ & Solid diffusivity \\
$D_e$ & Electrolyte diffusivity \\
$\epsilon_k$ & Electrolyte volume fraction, $k \in \{n,s,p\}$ \\
$t^+$ & Transference number \\
$a_k$ & Solid surface area density, $k \in \{n,s,p\}$ \\
$m_k$ & Reaction rate, $k \in \{n,s,p\}$ \\
$U_{k,ref}$ & Reference open circuit potential, $k \in \{n,s,p\}$ \\

$x$ & Cell coordinate location across cell \\
$z$ & Unitless dimension across electrode \\
$\beta$ & Jacobsen-West transfer function parameter \\
$S_{e,m}$ & Number of spacial electrolyte locations \\
$\mathscr{H}_{m}$ & Number of Hankel columns \\
$\mathscr{H}_{n}$ & Number of Hankel rows \\
$T_{len}$ & Transfer function sampling length \\
$F_s$ & Transfer function sampling frequency \\
$S_{s,m}$ & Number of spacial electrode locations \\
M & System order \\
$T_s$ & Final system sampling time \\

F & Faraday's constant \\
T & Cell Temperature \\
R & Universal gas constant \\
\hline

\end{tabularx}
\end{table}

\vspace{3em}
\section{\label{sec:Method}Methodology}
{This section discusses the methodology for the model development starting with the initial high-order model. A high-level derivation of the continuous-form transfer functions is discussed, followed by an introduction to the improved reduction method. Finally, the novel open-source LiiBRA.jl package is introduced.
\subsection{\label{sec:Continuum}Continuum Order Model}
This work starts with the Doyle-Fuller-Newman (DFN) continuum model as the high-order model for reduction. This model, first presented in two main publications \cite{doyle_modeling_1993, fuller_simulation_1994}, is a popular choice for continuum level electrochemical battery modelling due to its ability to capture multi-scale electrochemical processes within a lithium-ion cell. The DFN describes electrochemical electrodes of scale $\sim 100\si{\micro\metre}$ and active material particle size of scale $\sim 1\si{\micro\metre}$. These length scales are modelled one-dimensionally and coupled to produce a pseudo two-dimensional model space, often alternatively known as the "P2D" model. The geometry captured includes three domains: the positive electrode, the negative electrode, and the separator. A diagram describing this structure is presented in Figure \ref{Cell-Model-DFN} below.\\

\begin{figure}[!th] 
\centering
    \includegraphics[width=0.5\linewidth]{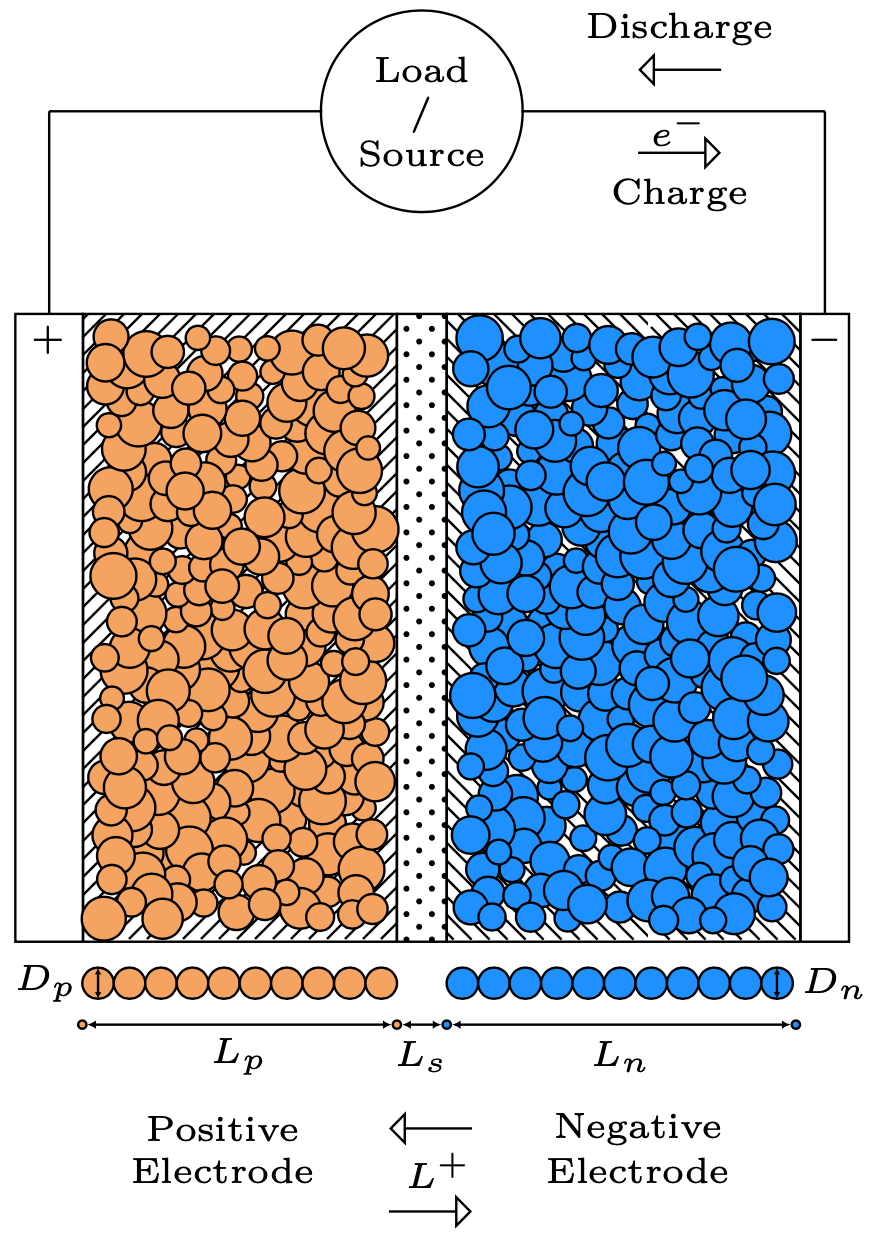}
    \caption{Doyle-Fuller-Newman model diagram with scale accurate sizing of electrodes, separator, and current collectors for an LG M50 cell.}
    \label{Cell-Model-DFN}
\end{figure}
\vspace{1em}

Diving further into this model, charge transfer reactions are distributed through electrode thickness $x \in [0, L_{tot}^*]$ with intercalated lithium diffused through the spherical domain $r \in [0, R_k^*] \ \text{where} \ k \in \{n,p\}$ and "$^*$" denotes variables of dimensionless domain. Porous electrode theory is utilised to capture mass and charge balance in the electrolyte and electrode domains and charge transfer kinetics. Through this theory, the porous electrodes are defined as a superposition of three states: electrolyte, electrochemically active material, and non-active material such as binders \cite{newman_porous-electrode_1975}. Concentration solution theory describes the ionic species transport in the electrolyte domain through relation of electrochemical potential gradient to the mass flux \cite{newman2012electrochemical}. The above, combined with the differentiated applied current density source term, gives an electrolyte mass balance equation (\ref{eq:1}) capturing the concentration evolution. Charge balance describes the current transferred from the porous electrode to the electrolyte with Ohm's law governing the charge conservation in the electrodes via equation (\ref{eq:6}). In the electrolyte the transport equation (\ref{eq:4}) accounts for ionic diffusion and migration. Coupled charge transfer kinetics in the electrode and electrolyte are captured via the Butler-Volmer equation in (\ref{eq:7}) by providing a relation between exchange current density and the potential difference between domains. Lastly, solid state (de)intercalation is represented through Fickian diffusion in equation (\ref{eq:3}). \\}

\begin{table}[h!]
\caption{Doyle-Fuller-Newman Governing Equations \cite{doyle_modeling_1993,fuller_simulation_1994, marquis_asymptotic_2019}}
\renewcommand{\arraystretch}{1.3}
\centering
\label{DFN_Equations}
\captionsetup{aboveskip=0pt,font=it}
\begin{tabularx}{0.9\textwidth}{>{\hsize=0.55\hsize\linewidth=\hsize}C 
    >{\hsize=0.62\hsize\linewidth=\hsize}C 
    >{\hsize=0.15\hsize\linewidth=\hsize}C @{}}
\toprule 
Governing Equation & Boundary Conditions & Equation Number \\
\toprule

\rowgroup{\textbf{Electrolyte Mass Conservation:}} \\
\multirow{2}{*}{$\frac{\partial (\epsilon_k c_{e,k}^*)}{\partial t^*} =  \frac{\partial N_{e,k}^*}{\partial x^*}+\frac{1}{F^*}\frac{\partial i_{e,k}^*}{\partial x^*},$} &
 $c_{s,n}^*|_{x^*=L_n^*}=c_{e,s}^*|_{x^*=L_n^*}$ & 
\multirow{2}{*}{\tagarray\label{eq:1}} \\ & 
$c_{s,p}^*|_{x^*=L^*-L_p^*}=c_{e,s}^*|_{x^*=L^*-L_p^*}$ \\
\multirow{3}{*}{$N_{e,k}^* = \epsilon_k^b D_e^*(c_{e,k}^*) \frac{\partial c_{e,k}^*}{\partial x^*} + \frac{t^+}{F^*} i_{e,k}^*$} & 
$N_{e,n}^*|_{x^*=0} = 0, N_{e,p}^*|_{x^*=L^*} = 0,$ \\ &
$N_{s,n}^*|_{x^*=L_n^*} = N_{e,s}^*|_{x^*=L_n^*},$&
\tagarray\label{eq:2} \\ &  
$N_{s,p}^*|_{x^*=L*-L_p^*} = N_{e,s}^*|_{x^*=L^*-L_p^*},$  \\
\midrule

\rowgroup{\textbf{Electrode Mass Conservation}} \\
\multirow{2}{*}{$\frac{\partial c_{s,k}^*}{\partial t^*} = \frac{1}{(r^*)^2} \frac{\partial}{\partial r^*} \bigg(D_{s,k}^*(r^*)^2 \frac{\partial c_{s,k}^*}{\partial r^*}\bigg)$} & $\frac{\partial c_{s,k}^*}{\partial r^*}|_{r^*=0} =  0,$ & \multirow{2}{*}{\tagarray\label{eq:3}}  \\ & $-D_{s,k}^* \frac{\partial c_{s,k}^*}{\partial r^*} = \frac{-j_{k}^*}{F^*}$  \\
\midrule

\rowgroup{\textbf{Charge Conservation}} \\
\multirow{5}{6cm}{$i_{e,k}^* = \epsilon_k^b \kappa_{e}^*(c_{e,k}^*) \bigg(-\frac{\partial \phi_{e,k}^*}{\partial x^*} +2(1-t^+)\frac{R^*T^*}{F^*}\frac{\partial}{\partial x^*}(\text{log}(c_{e,k}^*))\bigg),$} &
$i_{e,n}^*|_{x^*=0} = i_{e,p}^*|_{x^*=L^*} = 0,$ & 
\multirow{5}{*}{\tagarray\label{eq:4}} \\ & 
$\partial \phi_{e,n}^*|_{x^*=L_n^*} =\partial \phi_{e,s}^*|_{x^*=L_n^*} $  \\ &
$\partial i_{e,n}^*|_{x^*=L_n^*} =\partial i_{e,s}^*|_{x^*=L_n^*}  = I_{app}^*$  \\ &
$\partial \phi_{e,s}^*|_{x^*=L^*-L_p^*} =\partial \phi_{e,p}^*|_{x^*=L^*-L_p^*} $  \\ &
$\partial i_{e,p}^*|_{x^*=L^*-L_p^*} =\partial i_{e,p}^*|_{x^*=L^*-L_p^*} = I_{app}^*$  \\ 
$\frac{\partial i_{e,k}^*}{\partial x^*}= \begin{cases}  a_k^* j_k^*,& \text{k=n,p,}  \\ 0, & \text{n=s,}\\ \end{cases}$ & & \tagarray\label{eq:5} \\
$I_{app}^* - i_{e,k}^* = \sigma_{k}^*	\frac{\partial \phi_{s,k}^*}{\partial x^*}$ &  & \tagarray\label{eq:6} \\
\midrule

\rowgroup{\textbf{Charge Transfer Kinetics}} \\
$j_k^* = j_{0,k}^* \text{sinh} \bigg(\frac{F^*\eta_k^*}{2R^*T^*}\bigg),$ &  &  \tagarray\label{eq:7} \\
$j_{0,k}^* = m_k^*(c_{s,k}^*)^{1/2}(c_{s,k,max}^*-c_{s,k}^*)^{1/2}(c_{e,k}^*)^{1/2} ,$ &  & \tagarray\label{eq:8} \\
$\eta_{k}^* = \phi_{s,k}^* - \phi_{e,k}^*-U_{k}^*(c_{s,k}^*|_{r^*=R_k^*})$ &  & \tagarray\label{eq:9}  \\
	
\bottomrule
\end{tabularx}
\end{table}
\setcounter{equation}{9} 

\newpage
To simplify the spatial domains for the DFN, the respective dimensionless species lengths are transformed to,
\begin{equation}
	\gamma_n* = [0,L_n^*],
\end{equation}
\begin{equation}	
	\gamma_s^* = [L_n^*, L_n^*+L_s^*],
\end{equation}
\begin{equation}	
	\gamma_p^* = [L_n^*+L_s^*, L_t^*+L_p^*]
\end{equation}

Finally, the spacial variables are defined within the above domains and shown below. These variables are then numerically solved in the governing equations for the DFN.

\begin{table}[H]
\renewcommand{\arraystretch}{1.5}
\centering
\begin{tabularx}{\textwidth}{>{\hsize=0.33\hsize\linewidth=\hsize}C  
>{\hsize=0.33\hsize\linewidth=\hsize}C
>{\hsize=0.33\hsize\linewidth=\hsize}Q}

	&  $c_{e,n}*, \phi_{s,n}^*,\phi_{e,n}^*, i_{e,n}^*, N_{s,n}^*$ & $x \in [0,\gamma_n^*]$ \\

	& $c_{e,s}*, \phi_{e,s}^*, i_{e,s}^*, N_{e,s}^*$ & $x \in [\gamma_n*,\gamma_s^*]$\\

	&  $c_{e,p}*, \phi_{s,p}^*,\phi_{e,p}^*, i_{e,p}^*, N_{s,p}^*$ & $x \in [\gamma_s^*,\gamma_p^*]$\\
	
	& $c_{s,n}^*$ & $r^* \in [0,R_{n}^*]$, \ $x \in [0,\gamma_n^*]$\\
	
	& $c_{s,p}^*$ & $r^* \in [0,R_{p}^*]$, \ $x \in [\gamma_s^*,\gamma_p^*]$\\ 
	
\end{tabularx}
\end{table}

The above coupled system can be numerically solved through methods such as finite element\cite{smith_solid-state_2006}, finite difference\cite{subramanian_mathematical_2009}, orthogonal collocation\cite{bizeray_lithium-ion_2015,cai_lithium_2012,northrop_coordinate_2011}, and Chebyshev polynominals\cite{drummond_feedback_2020}; however, due to the model complexity of the coupled systems these methods are not feasible for implementation in current generation embedded automotive hardware. To achieve a real-time representation of the above model, this work forgoes these methods and investigates alternatives capable of capturing the resulting system dynamics. One such method, is the eigensystem realisation algorithm, which provides a data-driven approach to capturing system dynamics in a solvable state-space representation. This method requires observable state data to achieve stable, robust realisation; however, to achieve in-vehicle model model creation, the above coupled partial derivative system must be reduced to achieve fast solutions on available vehicle hardware. To accomplish this, a transfer function reduction is presented in the section below.\\

\subsection{\label{sec:Tfs}Derived Transfer Functions}
To achieve the required computational performance for in-vehicle model generation, the nonlinear governing equations shown above are reduced into low-order polynomial transfer functions. These transfer functions will be presented below, however, for a comprehensive derivation, the reader is pointed to Jacobsen and West\cite{jacobsen_diffusion_1995}, Smith et al.\cite{smith_control_2007}, and Lee et al.\cite{lee_one-dimensional_2012} The starting point for the derivation is linearising the Butler-Volmer equation produced in equation (\ref{linear_BV}). 

\begin{equation}
\label{linear_BV}
\widetilde{\Phi}_{s,e}(z,t) = FR_{tot}j(z,t)\bigg[\frac{\partial U_{ocp}}{\partial{C_{s,e}}} \biggl\lvert_{c_{s,0}}\bigg]\widetilde{C}_{s,e}(z,s)
\end{equation}

Next, a transfer function solution of the electrode surface concentration as shown in (\ref{eq:6}) has been presented by Jacobsen and West\cite{jacobsen_diffusion_1995} and is shown in (\ref{Conc_SmithWest}) below.

\begin{equation}
\label{Conc_SmithWest}
\frac{\widetilde{C}_{s,e}(z,s)}{J(z,s)} = \frac{R_s}{D_s} \left( \frac{\text{tanh}(\beta)}{\text{tanh}(\beta)-\beta} \right)
\end{equation}

where, $\widetilde{C}_{s,e}$ is the debiased surface concentration defined as $\widetilde{C}_{s,e} = C_{s,e}-C_{s,0}$, and $\beta = R_s \sqrt{\frac{s}{D_s}}$. Furthermore, by applying Laplace transformations to both the linearised Butler-Volmer equation and the solid phase charge conservation (\ref{eq:6}) while utilising the transfer function solution, the transfer function defining the electrode surface potential is presented in equation (\ref{surfpoteqn}).

\begin{equation}
\label{surfpoteqn}
	\begin{split}
	\frac{\widetilde{\Phi}_{s,e}(z,s)}{I_{app}(s)} = &\frac{L_n}{A \nu(s)\text{sinh}(\nu(s))} \Bigg( \frac{1}{\kappa^{\text{eff}}}\text{cosh}(\nu(s)z \\
	& +\frac{1}{\sigma^{\text{eff}}\text{cosh}(\nu(s)(z-1))}\Bigg) 
	\end{split}
\end{equation}

where $\nu(s)$ is a dimensionless variable defined as:

\begin{equation}
\nu(s) = L_n\sqrt{\frac{a_s(\frac{1}{\sigma^{\text{eff}}}+\frac{1}{\kappa^{\text{eff}}})}{R_{\text{tot}}+\Big[\frac{\partial U_{k,ref}}{\partial{C_{s,e}}}\Big] \frac{R_s}{F D_s}\Big(\frac{\text{tanh}(\beta)}{\text{tanh}(\beta)-\beta}\Big)}}
\end{equation}

The reaction flux transfer function can then be solved by combining the results obtained through the linearised Butler-Volmer and the surface potential in (\ref{surfpoteqn}). The final form is shown in equation \ref{fluxeqn} below.

\begin{equation}
\label{fluxeqn}
\begin{split}
\frac{J(z,s)}{I_{app}(s)} = &\bigg(\frac{\nu(s)}{a_sFL_nA(\kappa^{\text{eff}}+\sigma^{\text{eff}})}\bigg) \\ 
& +\bigg(\frac{\sigma^{\text{eff}}\text{cosh}(\nu(s)z)+\kappa^{\text{eff}}\text{cosh}(\nu(s)(z-1))}{\text{sinh}(\nu(s))}\bigg)
\end{split}
\end{equation}

Using the results from the previous steps, the electrode surface concentration transfer function can be derived as,

\begin{equation}
\label{Cse_eqn}
\begin{split}
\frac{\widetilde{C}_{s,e}(z,s)}{I_{\text{app}}(s)} = & \bigg(\frac{\nu(s)R_s(z-1)))}{a_sFL_nA(\kappa^{\text{eff}}+\sigma^{\text{eff}}))} \\ 
& \times \frac{(\sigma^{\text{eff}}\text{cosh}(\nu(s)\cdot z)+\kappa^{\text{eff}}\text{cosh}(\nu(s))}{\text{sinh}(\nu(s))} \\ 
&\times \frac{\text{tanh}(\beta)}{(\text{tanh}(\beta)-\beta)}\bigg)
\end{split}
\end{equation}

Starting with the solid phase charge conservation and integrating, the corresponding solid potential $\phi_s$ transfer function is,

\begin{equation}
\label{phis_eqn}
\begin{split}
\frac{{\phi}_{s}(z,s)}{I_{app}(s)} = &-\frac{L_n \kappa^{\text{eff}}(\text{cosh}((z-1)\nu(s)))}{A\sigma^{\text{eff}}(\kappa^{\text{eff}}+\sigma^{\text{eff}})\nu(s)\text{sinh}(\nu(s))}\\ 
& -\frac{L_n\sigma^{\text{eff}}(1-\text{cosh}(z\nu(s))+z\nu(s)\text{sinh}(\nu(s)))}{A\sigma^{\text{eff}}(\kappa^{\text{eff}}+\sigma^{\text{eff}})\nu(s)\text{sinh}(\nu(s))}\\ 
\end{split}
\end{equation}

The electrolyte potential can be likewise found through integration of electrolyte charge conservation equation (\ref{eq:3}) and combining with the previous reaction flux transfer function to obtain an ionic current representation. This process produces a two-term representation that is shown and further expanded on below.

\begin{equation}
\label{Electrolyte-Combination}
\frac{{\widetilde\phi}_{e}(z,s)}{I_{\text{app}}(s)}  = [{\widetilde\phi}_{e}(z,s)]_1 +[{\widetilde\phi}_{e}(z,s)]_2
\end{equation}

where the first term is obtained through the previously defined transfer functions as,
\begin{equation}
\label{phie1_eqn}
\begin{split}
[{\widetilde\phi}_{e}(z,s)]_1 = &-\frac{L_s}{A\kappa_{s}^{\text{eff}}}+\frac{L_n\bigg(\bigg(1-\frac{\sigma_n^{\text{eff}}}{\kappa_n^{\text{eff}}}\bigg)\text{tanh}\big(\frac{\nu_n(s)}{2}\big)-\nu_n(s)\bigg)}{A(\kappa_n^{\text{eff}}+\sigma_n^{\text{eff}})\nu_n(s)}\\ 
& -\frac{L_p\bigg(1+\frac{\sigma_p^{\text{eff}}}{\kappa_p^{\text{eff}}}\text{cosh}(\nu_p(s))\bigg)}{A(\kappa^{\text{eff}}+\sigma^{\text{eff}})\text{sinh}(\nu_p(s))\nu_p(s)}\\ 
&+\frac{L_p\text{cosh}\bigg(\frac{(L_n+L_s-x)\nu_p(s)}{L_p}\bigg)}{A(\kappa^{\text{eff}}+\sigma^{\text{eff}})\text{sinh}(\nu_p(s))\nu_p(s)}\\
&+\frac{L_p\bigg(\frac{\sigma_p^{\text{eff}}}{\kappa_p^{\text{eff}}}\text{cosh}\big(\frac{(L_t-x)\nu_p(s)}{L_p}\big)\bigg)}{A(\kappa^{\text{eff}}+\sigma^{\text{eff}})\text{sinh}(\nu_p(s))\nu_p(s)}\\
& + \frac{(L_n+L_s)-x}{A(\sigma_p^{\text{eff}}+\kappa_p^{\text{eff}})}\\
\end{split}
\end{equation}

and the second term is determined by the value of $c_e(x,t)$ and is shown as,

\begin{equation}
\label{phie2_eqn}
[\widetilde\phi_e(x,t)]_2 = \frac{2RT(1-t_+^0)}{F}\log\bigg(\frac{c_e(x,t)}{c_e(0,t)}\bigg)
\end{equation}

Finally, to acquire the electrolyte concentration transfer function the problem is split into homogenous and non-homogenous components. This allows for the homogenous component to obtain an orthonormal eigenfunction representation of the $\epsilon_e$ weighting function through separation of variables. The non-homogenous component performs a projection of the concentration function into $\epsilon_e$ to solve for Fourier coefficients. Finally, these are then used to derive the electrolyte concentration transfer function. This derivation is complex and the detail is not presented in this paper, but the final steps in the process give the negative electrode reaction flux as,

\begin{equation}
\label{jn_eqn}
\begin{split}
\frac{{j}^{\text{neg}}(s)}{I_{\text{app}}(s)} = &\frac{k_1(1-t_+^0)\hat{L}_n\text{sin}(\hat{L}_n)(\kappa_n^{\text{eff}}+\sigma_n^{\text{eff}}\text{cosh}(\nu_n(s)))\nu_n(s)}{AF(\kappa_n^{\text{eff}}+\sigma_n^{\text{eff}})(\hat{L}_n^{2}+\nu_n^2(s))\text{sinh}(\nu_n(s))} \\ 
&+\frac{k_1(1-t_+^0)\hat{L}_n\text{sin}(\hat{L}_n^)(\kappa_n^{\text{eff}}+\sigma_n^{\text{eff}})\nu_n^2(s)}{AF(\kappa_n^{\text{eff}}+\sigma_n^{\text{eff}})(\hat{L}_n^{2}+\nu_n^2(s))} \\ 
\end{split}
\end{equation}

where $\hat{L}_n = L_n\sqrt{\epsilon_n\lambda_k/D_n}$, and $\lambda_k$ are the eigenvalues obtained from the homogenous problem, which is numerically obtained through root finding. The corresponding positive electrode definition is then,\\

\begin{equation}
\label{jp_eqn}
\begin{split}
\frac{{j}^{\text{pos}}(s)}{I_{\text{app}}(s)} = &\frac{k_6(1-t_+^0)\hat{L}_p\text{cos}(\hat{L}_p)(\kappa_p^{\text{eff}}+\sigma_p^{\text{eff}}\text{cosh}(\nu_p(s)))\nu_p(s)}{AF(\kappa_p^{\text{eff}}+\sigma_p^{\text{eff}})(\hat{L}_p^{2}+\nu_p^2(s))\text{sinh}(\nu_p(s))} \\ 
& -\frac{k_5(1-t_+^0)\hat{L}_p\text{sin}\hat{(L}_p)(\kappa_p^{\text{eff}}+\sigma_p^{\text{eff}}\text{cosh}(\nu_p(s)))\nu_p(s)}{AF(\kappa_p^{\text{eff}}+\sigma_p^{\text{eff}})(\hat{L}_p^{2}+\nu_p^2(s))\text{sinh}(\nu_p(s))}\\ 
& +\frac{k_6(1-t_+^0)\hat{L}_p\text{cos}(\hat{L}_{ns})(\kappa_p^{\text{eff}}+\sigma_p^{\text{eff}}\text{cosh}(\nu_p(s)))\nu_p(s)}{AF(\kappa_p^{\text{eff}}+\sigma_p^{\text{eff}})(\hat{L}_p^{2}+\nu_p^2(s))\text{sinh}(\nu_p(s))}\\ 
& -\frac{k_5(1-t_+^0)\hat{L}_psin(\hat{L}_t)(\kappa_p^{\text{eff}}+\sigma_p^{\text{eff}}\text{cosh}(\nu_p(s)))\nu_p(s)}{AF(\kappa_p^{\text{eff}}+\sigma_p^{\text{eff}})(\hat{L}_p^{2}+\nu_p^2(s))\text{sinh}(\nu_p(s))}\\ 
& -\frac{k_5(1-t_+^0)\sigma_p^{\text{eff}}(\text{cos}(\hat{L}_{ns})\kappa_p^{\text{eff}}+\text{cos}(\hat{L}_s)\sigma_p^{\text{eff}})\nu_p^2(s)}{AF(\kappa_p^{\text{eff}}+\sigma_p^{\text{eff}})(\hat{L}_p^{2}+\nu_p^2(s))}\\
& -\frac{k_6(1-t_+^0)\sigma_p^{\text{eff}}(sin(\hat{L}_{ns})\kappa_p^{\text{eff}}+\text{sin}(\hat{L}_s)\sigma_p^{\text{eff}})\nu_p^2(s)}{AF(\kappa_p^{\text{eff}}+\sigma_p^{\text{eff}})(\hat{L}_p^{2}+\nu_p^2(s))}\\  
\end{split}
\end{equation}

with, $\hat{L}_p = L_p\sqrt{\epsilon_p\lambda_k/D_p}$, $\hat{L}_{ns}=L_ns\sqrt{\epsilon_p\lambda_k/D_p}$, and $\hat{L}_p = L_t\sqrt{\epsilon_p\lambda_k/D_p}$. As well, $k_1, \ k_3, \ k_4, \ k_5, \ \text{and} \ k_6$ are obtained by solving the system of equations below. 

\begin{equation*}
\Psi_n(x;\lambda) = k_1 cos\big(\sqrt{\frac{\lambda\epsilon_n}{D_n}}x\big)	 
\end{equation*}
\begin{equation*}
\Psi_m(x;\lambda) = k_3 cos\big(\sqrt{\frac{\lambda\epsilon_m}{D_m}}x\big) + k_4 cos\big(\sqrt{\frac{\lambda\epsilon_m}{D_m}}x\big) 
\end{equation*}
\begin{equation*}
\Psi_p(x;\lambda) = k_5 cos\big(\sqrt{\frac{\lambda\epsilon_p}{D_p}}x\big) + k_6 cos\big(\sqrt{\frac{\lambda\epsilon_p}{D_p}}x\big) 
\end{equation*}

where, $\Psi_k$ is the corresponding eigenfunction for the corresponding electrodes and separator. Finally, the electrolyte concentration is presented as,

\begin{equation}
\frac{C_{e,k}(x,s)}{I_{\text{app}}(s)} = \frac{1}{s+\lambda_k}\bigg[\frac{J_k^{neg}(s)}{\text{app}(s)}+\frac{J_k^{pos}(s)}{\text{app}(s)}\bigg]
\end{equation}

Combining the above individual transfer functions into a single input multiple output (SIMO) response array provides a single mathematical structure comprising of the continuous-time cell impulse response. Efficiently translating this formation into a state-space representation is the basis for the computationally informed discrete realisation algorithm defined in section \ref{sec:DRA} below.

\begin{equation}
\label{fulltransfer}
 Y(t) = \renewcommand*{\arraystretch}{1.8} \begin{Bmatrix}
			\frac{C_e(x,s)}{I_{app}(s)} \\
			\frac{\phi_e(x,s)}{I_{app}(s)} \\
			\frac{\tilde{C}_{s,e}(z,s)}{I_{app}(s)} \\
			\frac{\phi_{s}(z,s)}{I_{app}(s)} \\
			\frac{J(z,s)}{I_{app}(s)}  \\
			\end{Bmatrix}
\end{equation}
\vspace{1em}

\subsection{\label{sec:DRA}Computationally Informed Discrete Realisation Algorithm}
To reduce the presented Doyle-Fuller-Newman model, eigensystem realisation algorithms (ERA) are implemented to create real-time capable state space representations. This method utilises the sampled impulse response from the transfer functions derived in section \ref{sec:Tfs} above. The ERA provides a mathematical pathway to achieve the linear state-space realisation of the partial differential system of the form shown in equations (\ref{State Equation}) and (\ref{Output Equation}) below. 

\begin{equation}
\label{State Equation}
\dot{x}(t) = \textbf{A} x(t)+ \textbf{B} u(t)
\end{equation}
\begin{equation}
\label{Output Equation}
y(t) = \textbf{C} x(t) + \textbf{D} u(t)
\end{equation}

However, to deploy this system onto embedded hardware, a discrete-time realisation of equations (\ref{State Equation}) and (\ref{Output Equation}) is required. To achieve this, the discrete realisation algorithm (DRA) presented by Lee et al. \cite{lee_discrete-time_2012} allows for direct realisation of a discrete form state space model, as shown in equations (\ref{dState}) and (\ref{dOut}) below. In this work, a computationally informed discrete realisation algorithm is introduced, expanding on the conventional DRA.

\begin{equation}
\label{dState}
x[k+1] = \textbf{A} x[k]+ \textbf{B} u[k]
\end{equation}
\begin{equation}
\label{dOut}
y[k] = \textbf{C} x[k] + \textbf{D} u[k]
\end{equation}

The CI-DRA incorporates the zero-order hold methodology first presented in the conventional DRA, however, the bandwidth and sampling time is optimised to capture the system dynamics while providing fast solution times. First, the approximate discrete system response is introduced as, 

\begin{equation}
\label{approx_response}
G(z) \approx G(s)\mid_{s=\frac{2(z-1)}{T_i(z+1)}}
\end{equation}

where $T_i$ is the transfer function sampling period, and equivalent to the final system sampling period in the CI-DRA algorithm. This approximation is one of the key performance improvements in the CI-DRA and allows for further simplifications over the conventional DRA; however, care needs to be taken to ensure an adequate response is captured from the impulse-response. Next, by relating the discrete Fourier transformation of a sequence to its z-transform\cite{oppenheim2001discrete}, equation \ref{approx_response} can be defined as,

\begin{equation}
\label{approx_response_dft}
G_d[f] = G\bigg(\frac{2}{T_i}\frac{\exp(j2\pi f/N)-1}{\exp(j2\pi f/N)+1}\bigg)
\end{equation}

where $N$ is the number of points for the transfer function response and is sized to be memory efficient instead of computationally efficient in the conventional DRA. Through this sizing, the CI-DRA provides additional memory optimisation as $N$ impacts the full realisation process. Finally, due to constraining $T_s=T_i$, the CI-DRA can forgo the interpolation steps required in the conventional DRA. This removal provides a substantial performance improvement for realisation with large values of $N$. Additionally, this allows for direct manipulation of the discretely sampled transfer function responses, without additional cumulation steps that are required in the conventional DRA. Continuing the realisation process, the Ho-Kalman\cite{ho_effective_1966} algorithm is utilised to generate the state-space representation. This is completed by understanding the resultant discrete response comprises the Markov parameters, $G_{(t)}$, for the system, this response can be shown in the following form,

\begin{equation}
\label{markov}
\begin{split}
G_{(t)}=
\begin{cases}
      D & t=0  \\
      CA^{t-1}B & t=1,2,3,..\\
    \end{cases}       
\end{split}
\end{equation}
\vspace{0.5em}

where D is gathered from the system at time step zero. The resultant transfer function response can be formulated into a block Hankel matrix, (\ref{BlockHankel}), of Markov parameters. This block Hankel, $\mathcal{H}_{k,m}$, has indices corresponding to a subset domain of the discrete-time impulse response.\\ 

\begin{equation}
\label{BlockHankel}
\mathcal{H}_{k,m}= 
\begin{pmatrix}
G_1& G_2 & G_3 & \cdots  & G_{m}\\
G_2 & G_3 & G_4 & \cdots & G_{m+1}\\
G_3 & G_4 & G_5 & \cdots & G_{m+1}\\
\vdots & \vdots & \vdots &\ddots & G_{m+l}\\
G_{k} & G_{k+1} & G_{k+2} & \cdots & G_{m+k-1} \\
\end{pmatrix}
\end{equation}
\vspace{0.5em}

An additional feature of the block Hankel matrix is its relation to the controllability and observability matrices as shown in equation \ref{hankel_form}. The observability $(\mathscr{O}) \ \text{and the controllability} \ (\mathscr{C})$ matrices are defined in equations \ref{observe_matrix} and \ref{control_matrix} below. By exploiting this relation and factoring $\mathscr{H}_{k,m}$ into the two matrices it is possible to obtain the A matrix. 

\begin{equation}
\label{hankel_form}
 \mathcal{H}_{k,m} = \mathscr{O}_{k}\mathscr{C}_{m}
\end{equation}

\begin{equation}
\label{observe_matrix}
\mathscr{O} = \begin{bmatrix}
 			C \\
			CA \\
			CA^2 \\
			\vdots\\
			CA^{k-1}
			\end{bmatrix}
\end{equation}

\begin{equation}
\label{control_matrix}
\mathscr{C} = \begin{bmatrix}
 			B\quad AB\quad A^2B\quad \cdots\quad A^{m-1}B
			\end{bmatrix}
\end{equation}

To accomplish this factoring, singular value decomposition (SVD) provides the mechanism to reduce the block Hankel through truncation of the system order. The truncated SVD is shown in equation \ref{SVD} below, where $\Sigma_s$ captures the highest order singular values of the block Hankel in descending order while $\Sigma_n$ captures the remaining values and is approximately zero. Selection of the size of $\Sigma_s$ is dependent on the numerical performance requirements and available computation power for decomposition.

\begin{equation}
\label{SVD}
 \mathcal{H}_{k,m} = \begin{bmatrix}
 				U_s & U_n
			   \end{bmatrix} 
			   \begin{bmatrix}
 				\Sigma_s & 0 \\
				0 & \Sigma_n
			   \end{bmatrix} 
			   \begin{bmatrix}
 				V_s^T \\ 
				V_n^T
			   \end{bmatrix}    
			   = U_s\Sigma_sV_s^T
\end{equation}
\vspace{0.25em}

where the structure for the observability and controllability matrices is then:

\begin{equation}
\label{Observe_SVD}
 \mathcal{O}_k = U_s\Sigma_s^{1/2}T
\end{equation}

\begin{equation}
\label{Control_SVD}
\mathcal{C}_l = T^{-1}\Sigma_l^{1/2}V_l^T
\end{equation}

From these formed matrices, we can exploit the structure of the Markov parameters and obtain the resulting state-space representation as:

\begin{equation}
 A = \mathscr{O}_{k}^\dag \mathscr{H}_{k,m+1} \mathscr{C}_m^\dag
\end{equation}

\begin{equation}
B=\mathscr{C}_l[1{:}n,1{:}m]
\end{equation}

\begin{equation}
 C = \mathscr{O}_k[1{:}p,1{:}n]
 \end{equation}
 
where $n$ and $p$ are the state-space output size and input size respectively, and $\mathscr{H}_{k,m+1}$ denotes a single indices forward shifted of the block Hankel matrix. With the above equations, the realisation process is completed and a linear system model of state-space form as shown previously in equation \ref{dState} and \ref{dOut} is acquired.\\ 

In this work, the conventional discrete realisation algorithm is expanded on, with the aim to improve computational performance. To accomplish this, the conventional method was profiled and computational bottlenecks were noted with mitigations implemented. This computationally informed discrete realisation algorithm (CI-DRA) removes the interpolation requirements when the transfer function sampling time and final system sampling time are equivalent. Furthermore, if the final system sampling time is a subset of the transfer function sampling time, the scalar multiple of the subset is used for the remaining steps in the DRA. In simplistic terms if $F_s$ is devisable by $T_s$ the indices used are a subset of that division. Next, the CI-DRA aims to maximise in-place functional operations to minimise memory overhead in the realisation steps. The improved discrete realisation algorithm structure is shown in the list below. \\

Summarry of the CI-DRA:
\begin{enumerate}
\item Compute the continuous form frequency response $G(f)$ from derived transfer functions and convert to discrete-time impulse response via inverse fast Fourier transformation.
\item Form the block Hankel matrix from the discrete-time impulse response and compute the in-place singular value decomposition.
\item Perform an in-place operation on the block Hankel matrix to obtain the time-shifted block Hankel matrix.
\item Form linear state-space arrays from Ho-Kalman algorithm with unstable poles replaced by their magnitudes, and oscillating poles replace by reciprocals.
\end{enumerate}
\vspace{1em}

\subsection{\label{sec:Comp_Imp}Computational Implementation}
LiiBRA.jl, a Julia\cite{Julia-2017} based package, has been created by the authors for fast computational implementation of the above computationally informed discrete realisation algorithm. This package aims to improve on previous implementations of the eigensystem realisation algorithm by improving computation performance while maintaining fidelity. Key improvements include performant truncated SVD support, large array storage optimisation, and performance benefits acquired from the Julia language largely due to the bottleneck from block Hankel matrix formation. Julia provides a high performance dynamic typeset, with just-in-time compilation and multiple dispatch capabilities. These features provide an effective computational language for scientific computing, while providing a modern syntax. The open-source code repository for LiiBRA.jl can be found at: \url{https://github.com/BradyPlanden/LiiBRA.jl}.\\

\begin{figure}[H] 
\centering
    \includegraphics[width=0.85\linewidth]{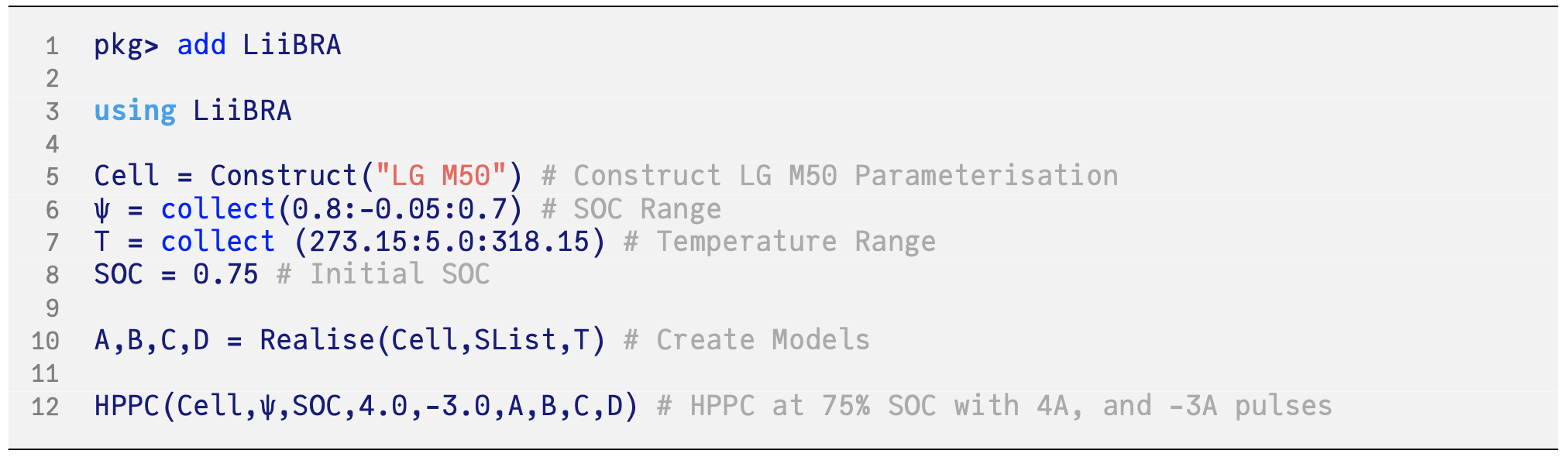}
    \caption{Example usage of LiiBRA.jl, providing a simple package for creating and simulating reduced-order models}
    \label{Example-Script}
\end{figure}

LiiBRA.jl provides a framework to define, create, and simulate lithium-ion cell realisation algorithms. This is accomplished with a fundamental aim towards reducing the computational requirements to generate embeddable state-space models. To easily generate a state-space model, a package was developed that could provide simple function calls to users. This has been accomplished and is presented in the code example shown in Figure \ref{Example-Script}. A mutable struct is utilised for storage of cell parameterisation, as well as the realisation algorithm variables. This struct is exported throughout the package and provides a method to minimise memory allocation through in-place function handles and storage. This helped with the aim of generating an open-source package for improved collaboration and advancement of the real-time battery modelling field. \\

\vspace{1em}

The high-level structure of LiiBRA.jl is shown in Figure \ref{LiiBRA_Arch} below. The package dependencies are shown and offer improved code reusability while minimising the package size of LiiBRA.jl. Through distributing the code base and utilising Julia's open-source packages, LiiBRA.jl can be modular and flexible  providing improved algorithm selection for compatibility and performance. These dependencies include TSVD.jl\cite{larsen_lanczos_1998} for the truncated SVD, Interpolations.jl\cite{dierckx1995curve} for spline fitting of the resultant impulse response, and FFTW.jl\cite{FFTW.jl-2005}, providing interface for inverse fast Fourier transforms. \\

\vspace{1em}
\begin{figure}[H] 
\centering
    \includegraphics[width=1\linewidth]{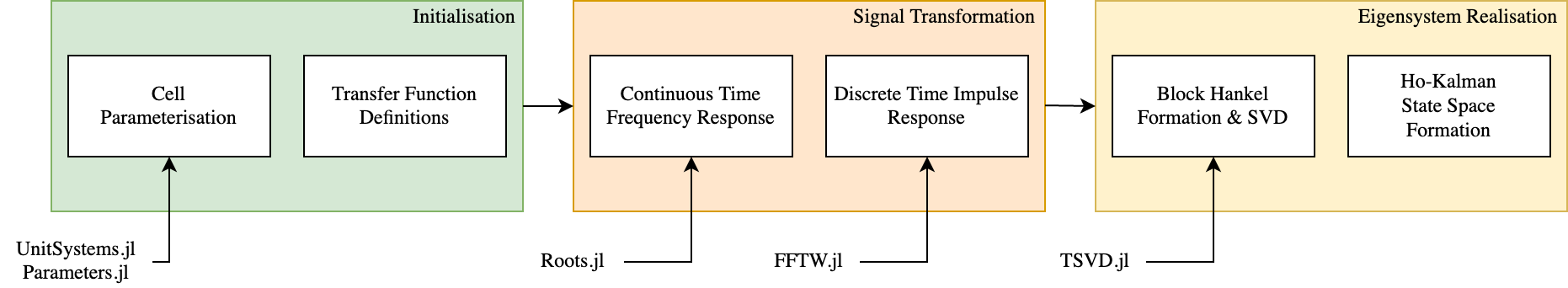}
    \caption{High-level flowchart of LiiBRA.jl's implementation of CI-DRA with package dependencies listed}
    \label{LiiBRA_Arch} 
\end{figure}
\vspace{1em}

A main achievement for LiiBRA.jl has been implementation of the computationally informed discrete realisation algorithms. To provide stable operation, LiiBRA.jl implements the conventional DRA as a fallback method when the conditions required for the CI-DRA are not met. This provides an easy interface for model creation, with feedback provided to end-users on the computational method being utilised. The transfer function sampling length has also been investigated, and computational improvements have been found by minimising the memory requirements for this variable. In previous implementations\cite{rodriguez_comparing_2019} of the DRA, this value has been rounded to the nearest power of two, whereas LiiBRA.jl avoids this addition memory requirement.\\

\section{\label{sec:Results}Results}
{LiiBRA.jl has been implemented on a 2019 Macbook Pro 13" Intel i5 with results gathered to investigate the performance capabilities, as well as experimental validation and numerical verification. First, the singular value decomposition has been found to have a large impact on the total numerical solution time for the conventional discrete realisation algorithm. To achieve the performance aims in this work, optimising the truncated singular value decomposition is required.  Three algorithms have been investigate due to their high computational efficiency: Arpack.jl\cite{Lehoucq97arpackusers},  PROPACK.jl\cite{dominique_2021_5090075}, and TSVD.jl\cite{larsen_lanczos_1998}.\\ 

Arpack.jl is a Julia to Fortran wrapper package of the implicitly restarted Arnoldi method\cite{arnoldi1951principle}, reducing to the implicitly restarted Lanczos method for symmetric matrices. PROPACK.jl is likewise a Julia to Fortan wrapper of the Fortran PROPACK software, initially developed by R.M. Larsen\cite{larsen_lanczos_1998}. This package implements the Lanczos bidiagonalization method with partial reorthogonalization and implicit restart in which it acts directly on the system matrix without forming the equivalent system in memory. Similarly, TSVD.jl implements the Lanczos bidiagonalization method with partial reorthogonalization; however, it is implemented directly into Julia.\\

These three algorithms have been compared for use with LiiBRA.jl, with the results presented in Figure \ref{SVD_Comp} below. TSVD.jl provides the lowest solution time across the presented range of block Hankel matrices, with PROPACK.jl and Arpack.jl following respectively. It can be seen that PROPACK.jl uses significantly more memory with the difference between TSVD.jl and Arpack.jl being negligible. TSVD.jl has been chosen as the best algorithm as it can provide a large range of solutions regarding Hankel size without having to compromise on speed. Selection between these three algorithms provides a control mechanism for optimal solutions for varying hardware capabilities. \\}

\vspace{1em}
\begin{figure}[!ht]
\centering
    \begin{tikzpicture}
      \begin{axis}[ 
      width=0.85\linewidth,
      line width=0.5,
      grid=major, 
      tick label style={font=\normalsize},
      label style={font=\small},
      grid style={white},
      xlabel={Hankel Size},
      ylabel={Computation Time (s)},
      legend style={at={(0.3,0.95)}, anchor=north east,  draw=none, fill=none},
      ytick={0,1,...,12},
      xtick={500,1000,...,5500},
      ymin = -.5,
      ymax = 12,
      xmin = 1350,
      xmax = 5150,
      axis y line*=left,
  ]
         \addplot[dashed] coordinates {(1500,0)};
	 \label{Memory1}
	
   \addplot+[color=DodgerBlue, only marks, mark=triangle*, mark options={solid, fill=DodgerBlue}] coordinates
      {(1000, 0.2393) (1500,0.6428) (2000,1.162) (2500,1.978) (3000,2.377) (3500,3.204) (4000,4.255) (4500, 5.27) (5000,6.519)};
      \label{Pro}
      
    \addplot+[color=SandyBrown, only marks, mark=square*, mark options={solid, fill=SandyBrown}] coordinates
      {(1000, 0.1353) (1500,0.3756) (2000,0.6831) (2500,1.181) (3000,1.419) (3500,1.906) (4000,2.524) (4500, 3.133) (5000,3.891)};
      \label{TSVD}

    \addplot+[color=DarkSeaGreen, only marks, mark=*, mark options={solid, fill=DarkSeaGreen}] coordinates
     {(1000, 0.2739) (1500,0.7783) (2000,1.411) (2500,2.47) (3000,2.976) (3500,3.987) (4000,5.317) (4500,6.552) (5000,8.154)};
     \label{Ar}  

\end{axis}

\node [draw,fill=white] at (rel axis cs: 0.05,0.90) {\shortstack[l]{
\ref{Pro} PROPACK \\
\ref{Ar} Arpack \\
\ref{TSVD} TSVD}};

  \begin{axis}[ 
      width=0.85\linewidth,
      line width=0.5,
      grid=major, 
      tick label style={font=\normalsize},
      legend style={nodes={scale=0.6, transform shape}},
      label style={font=\small},
      axis y line*=right,
      axis x line=none,
      grid style={white},
      xlabel={Hankel Size},
      ylabel={Memory (GiB)},
      ytick={0,1,...,11},
      xtick={500,1000,...,5500},
      ymin = -0.25,
      ymax = 6,
      xmin = 1350,
      xmax = 5150,
  ]
  
         \addplot+[color=DodgerBlue, dashed, mark=triangle*, mark options={solid, fill=DodgerBlue}] coordinates
      {(1000,0.2284) (1500,0.5103) (2000,0.9039) (2500,1.4100) (3000,2.0300) (3500,2.7600) (4000,3.6000) (4500, 4.5500) (5000,5.6100)};
      \label{ProMem}
      
               \addplot+[color=SandyBrown, dashed, mark=square*, mark options={solid, fill=SandyBrown}] coordinates
      {(1000,0.0063) (1500,0.0094) (2000,0.0125) (2500,0.0156) (3000,0.0187) (3500,0.0218) (4000,0.0249) (4500, 0.0280) (5000,0.0311)};
      \label{TSVDMem}

              \addplot+[color=DarkSeaGreen, dashed, mark=*, mark options={solid, fill=DarkSeaGreen}] coordinates
     {(1000,0.0033) (1500,0.0049) (2000,0.0065) (2500,0.0081) (3000,0.0097) (3500,0.0114) (4000,0.0130) (4500,0.0146) (5000,0.0162)};
     \label{ArMem}
         
 \end{axis}

\node [draw,fill=white] at (rel axis cs: 0.05,0.8) {\shortstack[l]{
\ref{Memory1} Memory}};

\end{tikzpicture}
\caption{Computation results of PROPACK.jl, TSVD.jl, and Arpack.jl completing SVD of varying block Hankel sizes}
\label{SVD_Comp}
\end{figure}
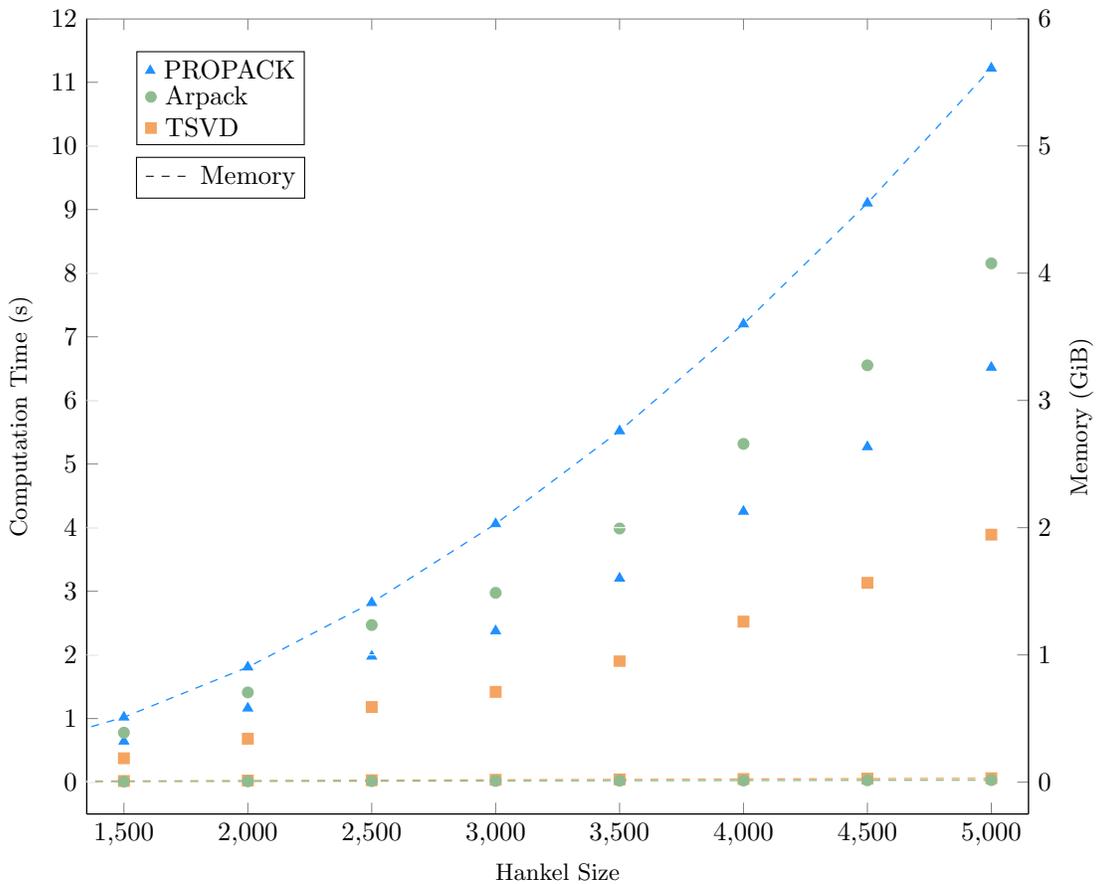
\vspace{1em}

\subsection{Computational Sensitivity}
{A numerical sensitivity analysis was also completed for the initialisation variables. The variables and the tested ranges are listed in Table \ref{tab:RA_Params}. Default values were selected as stable initial values to avoid numerical errors; however, are not necessarily the recommended final solution. The minimum and maximum of each range was tested to determine each variable's sensitivity on the resultant computational time. This analysis was completed by varying each variable individually while maintaining the remaining ones at the defined default parameters. The standard benchmarking package provided by Julia, BenchmarkingTools.jl\cite{BenchmarkTools.jl-2016}, was utilised to obtain the relevant statistical data. For this work, the minimum number of simulations for each variable set was selected at six to constrain the total number of simulations while reducing the effect of numerical jitter on the analysis. The median computation time for each variable is shown in Figure \ref{Num_Sens}. \\}

\vspace{1em}
\begin{table}[h!]
\caption{Default values and ranges for LiiBRA.jl sensitivity analysis of initialisation variables}
\centering
\label{tab:RA_Params}
\begin{tabular}{ccccc}
Variable & Definition & Default & Range\\ \hline
$S_{e,m}$ & Number of spacial particles in electrolyte & 6 & 4 - 8\\
$S_{s,m}$ & Number of spacial particles in electrodes & 4 & 2 - 6\\
$\mathscr{H}_{m}$ &  Number of columns included in Hankel matrices & 2500 & 1500 - 3500 \\
$\mathscr{H}_{n}$ & Number of rows included in Hankel matrices & 2500 & 1500 - 3500\\
$T_{len}$ & Length of transfer function sampling time [hr] & 4.5 & 1.0 - 8.0 \\
$F_s$ / $T_s$ & System sampling frequencies [Hz] & 4 & 2 - 6 \\
M & System order & 8 & 4 - 12\\
\end{tabular}
\end{table}
\vspace{1em}

{This analysis provides insight towards a minimal configuration of the package for improved computational time. The number of particles in the electrolyte $(S_{e,m})$, and transfer function sampling length $(T_{len})$ have the lowest sensitivities, and thus should be selected based on model fidelity. The model order (M), number of particles in the electrode$(S_{s,m})$, and block Hankel dimensions $(\mathscr{H}_{m}, \mathscr{H}_{n})$ have large impacts on the total computational time and these variables should be selected based on a compromise between model performance and solution time requirements. For this analysis, the transfer function sampling frequency $(F_s)$ and final system sampling time $(T_s)$ are modified together, as this provides a stable solution for investigating the improved realisation algorithm.\\
}

\vspace{1em}
\pgfplotstableread[col sep=comma]{
Parameter,Lb,Ub
$\mathscr{H}_m$ ,4.09,11.74
$\mathscr{H}_n$ ,4.1,11.70
$S_{s,m}$ ,5.35,10.26
M,5.46,9.54
$S_{e,m}$ ,6.9,8.42
$T_{len}$,7.37,7.76
$F_s$ / $T_s$,7.67,8.1
}\loadedtable

\newlength{\dy}\setlength{\dy}{\baselineskip}
\newlength{\dx}\setlength{\dx}{2em}
\newlength{\temp}

\pgfplotstablegetrowsof{\loadedtable}
\pgfmathparse{\pgfplotsretval-1}
\edef\rows{\pgfmathresult}
\setlength{\temp}{0pt}
\foreach \y in {0,1,...,\rows}{%
  \pgfplotstablegetelem{\y}{Lb}\of\loadedtable
  \global\advance\temp by \pgfplotsretval pt
  \pgfplotstablegetelem{\y}{Ub}\of\loadedtable
  \global\advance\temp by \pgfplotsretval pt
}
\pgfmathparse{0.01\temp}
\edef\total{\pgfmathresult}

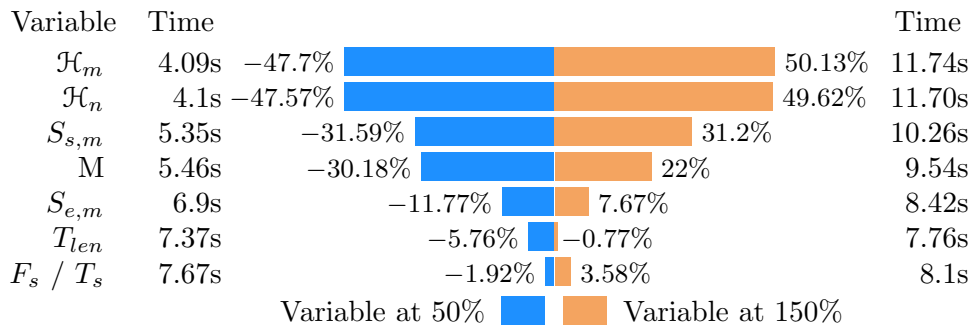
\begin{figure}[h!]
\centering
\resizebox{0.8\columnwidth}{!}{
\noindent\begin{tikzpicture}

\node at (-12.75em,1.25\dy) {Time};
\node at (12.75em,1.25\dy) {Time};
\node at (-16.75em,1.25\dy) {Variable};

\foreach \y in {0,1,...,\rows}{%
  \pgfplotstablegetelem{\y}{Parameter}\of\loadedtable
  \node[left] at (-15em,-\y\dy) {\strut\pgfplotsretval};
  \pgfplotstablegetelem{\y}{Lb}\of\loadedtable
  \node[left] at (-11em,-\y\dy) {\strut\pgfplotsretval s};
  \pgfmathparse{((\pgfplotsretval-7.82)/7.82*100)}
  \node[left,fill=DodgerBlue,text width=(0.075*abs(\pgfplotsretval-7.82)/7.82*100)*\dx,text height=.8\dy,inner sep=0]
    at (0,-\y\dy+0.1\dy) {};
    \pgfmathprintnumberto[fixed,precision=2]{\pgfmathresult}{\per}
    \pgfmathparse{(abs(1-\pgfplotsretval/7.82)*100))}
  \node[left] at (-\pgfmathresult\dx*0.075,-\y\dy) {\small\strut\per\%};
  
  \pgfplotstablegetelem{\y}{Ub}\of\loadedtable
  \node[left] at (14.5em,-\y\dy) {\strut\pgfplotsretval s};
    \pgfmathparse{(abs(\pgfplotsretval-7.82)/7.82*100)}
  \node[right,fill=SandyBrown,text width=(0.075*abs(\pgfplotsretval-7.82)/7.82*100)*\dx,text height=.8\dy,inner sep=0]
    at (0,-\y\dy+0.1\dy) {};
      \pgfmathparse{(abs(\pgfplotsretval)/7.82*100)-100}
      \pgfmathprintnumberto[fixed,precision=2]{\pgfmathresult}{\per}
        \pgfmathparse{(abs(\pgfplotsretval)/7.82*100)-100}
  \node[right] at (\pgfmathresult\dx*0.075,-\y\dy) {\small\strut\per\%};
}
\node[left] at (-2em,-\rows\dy-1\dy) {Variable at 50\%};
\node[left,fill=DodgerBlue,text width=1.5em,text height=.8\dy,inner sep=0]
  at (-1mm,-\rows\dy-1\dy) {};
\node[right,fill=SandyBrown,text width=1.5em,text height=.8\dy,inner sep=0]
  at (+1mm,-\rows\dy-1\dy) {};
\node[right] at (2em,-\rows\dy-1\dy) {Variable at $150\%$};

\end{tikzpicture}
}
\caption{CI-DRA numerical sensitivity results for simulating each variable at lower bound (50\% default) and higher bound (150\% default)}
\label{Num_Sens}
\end{figure}
\vspace{1em}

\subsection{Numerical Verification}
{For validation of the presented modelling framework, a worldwide harmonised light vehicle test procedure (WLTP)\cite{WLTP_2021} has been implemented, and is used as a representative operational use case for the generated reduced order models. This drive cycle was created from specifications provided for a Tesla Model 3, which are given in Table \ref{WLTP Drive-cycle} in the Appendix A below. The resultant specifications were utilised to generate the single-cell scaled power cycle as shown in the supplemental material, for a pack designed for LG Chem. M50 cells. Variables presented with an asterisk were used for calibration of the predictive model and are not aimed to be exact representations of the vehicle.\\
}

\begin{figure}[H]
\centering
\begin{tikzpicture}[spy using outlines=
	{rectangle, magnification=2, connect spies}]
	
	\begin{axis}[no markers,
	width=0.85\linewidth,
  	line width=0.5,
  	grid=major, 
  	tick label style={font=\normalsize},
	xmin=0,xmax=1800,
	grid style={white},
	ylabel style = {align=center},
        ylabel={Negative Electrode Concentration $\frac{mol}{m^3}$},
        legend style={at={(0.95,0.95)},anchor=north east,  draw=none, fill=none},
        xlabel=Time (s),
        tick label style={font=\normalsize},
        each nth point=4, filter discard warning=false, unbounded coords=discard]

		\addplot [color=DodgerBlue] table [x=PyBaMM_T, y=PyBaMM_Cn_20_20, col sep=comma] {WLTP.csv};
		\addplot [color=SandyBrown]table[x=LiiBRA_T, y=LiiBRA_Cn_2, col sep=comma] {WLTP.csv};

		\addlegendimage{/pgfplots/refstyle=plot2}\addlegendentry{PyBaMM}
		\addlegendimage{/pgfplots/refstyle=plot1}\addlegendentry{LiiBRA.jl}

  \coordinate (spypoint) at (axis cs:1100,21450);
  \coordinate (magnifyglass) at (axis cs:400,20750);
\end{axis}

\spy [black, size=3cm] on (spypoint)
   in node[fill=white] at (magnifyglass);
\end{tikzpicture}
\caption{Comparison of LiiBRA.jl's negative electrode concentration to the full order implementation in PyBAMM, WLTP 3B at 75\% SOC and 25\degree C}
\label{Cs-Neg-Comparision}
\end{figure}
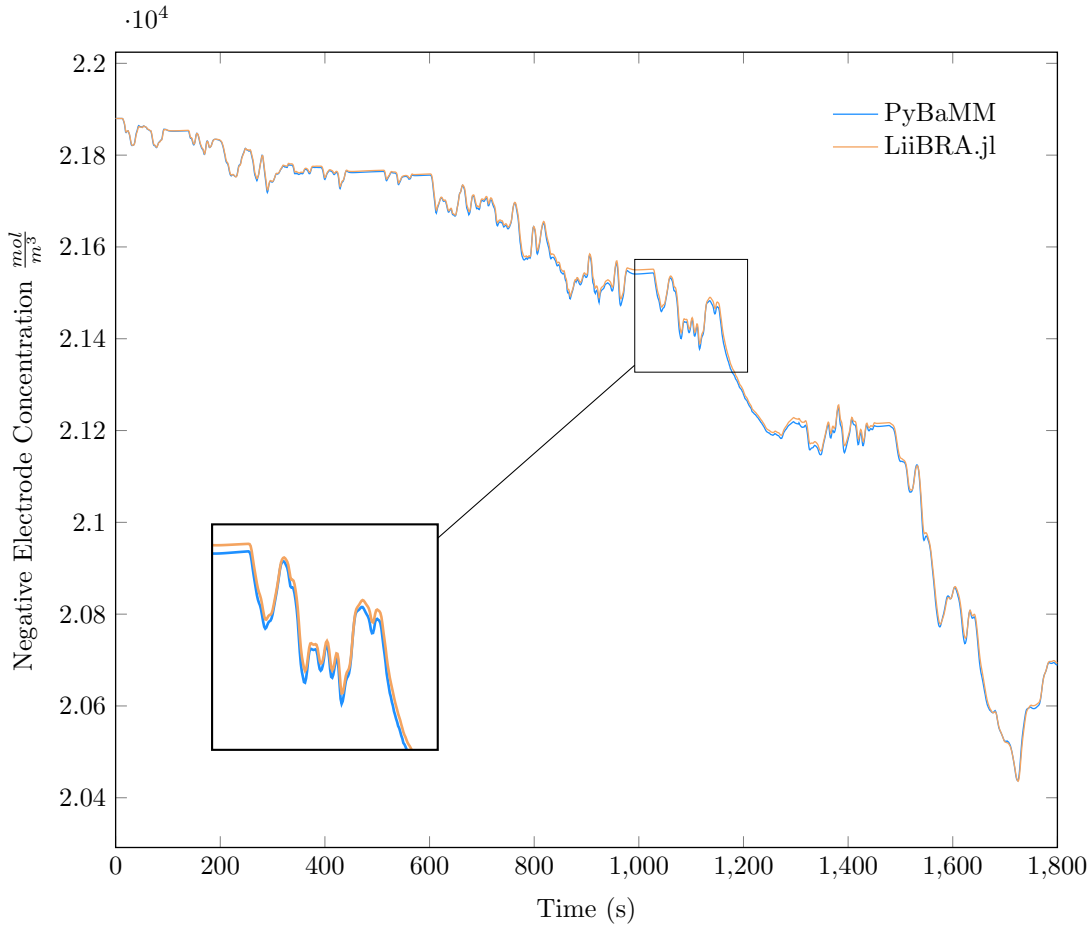

{LiiBRA.jl was then parameterised with the electrochemical characterisation presented by Chen et al. and was utilised to generate models as well as simulate the WLTP drive cycle. Additionally, the open-source python battery mathematical modelling package\cite{Sulzer_Python_Battery_Mathematical_2021} (PyBaMM) was utilised to solve the continuum order model with the identical parameterisation. Figure \ref{Cs-Neg-Comparision} outlines the predicted negative electrode concentration for both methods when initialised at the experimentally aligned 75\% state of charge value. A root-mean-square deviation between the two of 6.46 $mol/m^3$ and an absolute maximum deviation of 21.43 $mol/m^3$ was observed. For this verification, the block Hankel was sized at 2500 by 2500 elements, with the transfer function and final system sampling time set to 4 Hz, a system order of 6, and a transfer function sampling length of 4.5 hours. Total model creation time for this comparison was 21.75 seconds comprising of five SOC points from 100\% to 0\% for interpolation.  Expanding into full model formation, with a 25\% SOC increments and a temperature coverage from 5\degree C to 55\degree C in 10\degree C increments results in a total formation time of 130.5 seconds for LiiBRA.jl (x86) and 437.1 seconds for Matlab. \\
}

\vspace{1em}
\vspace{1em}

{Further performance investigations were performed against the conventional DRA via a previously presented Matlab implementation\cite{hu_comparative_2012, plett_battery_2015}. Both implementations were parameterised based on the default values in Table \ref{tab:RA_Params}, with computational timings recorded and presented in Figure \ref{Numeric_Comp} below. This comparison showcases the performance benefits of LiiBRA.jl for fast realisation algorithm computation through the improvements discussed in the above section.
This computational comparison results with a mean computational improvement in LiiBRA.jl's implementation of 2.73 times faster over the Matlab implementation. These results showcase the viability for in-vehicle model creation, and opens up physics-based model modifications over the life-time of the battery pack. \\}

{A key aspect for this work is in-vehicle model creation, both for ease of deployment and numerical performance. This capability provides an important mechanism for parameterisation modifications, degradation improvements, and transfer function improvements in-vehicle. Additionally, this mechanism allows for further cell characterisation to be completed off vehicle, with a final parameterisation to be provided during a standard charging operation, or when the vehicle is not in use. To investigate LiiBRA.jl's capabilities, the package was compiled onto a Qualcomm Snapdragon 845 ARM system-on-chip running Ubuntu 18.04.5 LTS operating system. Julia version 1.7.1 was used for this investigation, with results shown in Figure \ref{Numeric_Comp} below and showcase a capable solution to physics-based real-time model creation in-vehicle. For an equivalent parameterisation shown in Figure \ref{Cs-Neg-Comparision} above, the ARM based implementation can be completed in 5.5 seconds.\\}
\vspace{1em}

\begin{figure}[h!]
\centering
\begin{tikzpicture}
  \begin{axis}[ 
  width=0.82\linewidth,
  line width=0.5,
  grid=major, 
  tick label style={font=\normalsize},
  legend style={nodes={scale=0.6, transform shape}},
  label style={font=\small},
  grid style={white},
  xlabel={Hankel Size},
  ylabel={Computation Time (s)},
  legend style={at={(0.25,0.95)}, anchor=north east, draw=none, fill=none},
  ytick={0,4,...,46},
  xtick={500,1000,...,4000},
  ymin = 0,
  xmin = 250,
  xmax = 4250,
  ]
  
    \addplot+[color=SandyBrown, dashed, mark=x, mark options={solid, fill=SandyBrown}]coordinates
      {(500,1.72) (1000,2.18) (1500,2.90) (2000,3.89) (2500,5.04) (3000,6.63) (3500,8.34) (4000,10.36)};
      \addlegendentry{LiiBRA.jl}

           \addplot+[color=DodgerBlue, dashed, mark=square*, mark options={solid, fill=DodgerBlue}]  coordinates
     {(500,2.99) (1000,3.56) (1500,4.51) (2000,5.89) (2500,7.48) (3000,9.6) (3500,12.21)};
     \addlegendentry{LiiBRA.jl (ARM)}
     
         \addplot+[color=SeaGreen, dashed, mark=triangle*, mark options={solid, fill=SeaGreen}] coordinates
     {(500,2.463) (1000,4.08) (1500,6.93) (2000,10.87) (2500,15.49) (3000,21.6) (3500,28.74) (4000,36.77)};
     \addlegendentry{MATLAB}
     
  \end{axis}
\end{tikzpicture}
\caption{Computation results for CI-DRA using LiiBRA.jl (x86, ARM) and the conventional DRA using Matlab (x86) for varying block Hankel sizes}
\label{Numeric_Comp}
\end{figure}
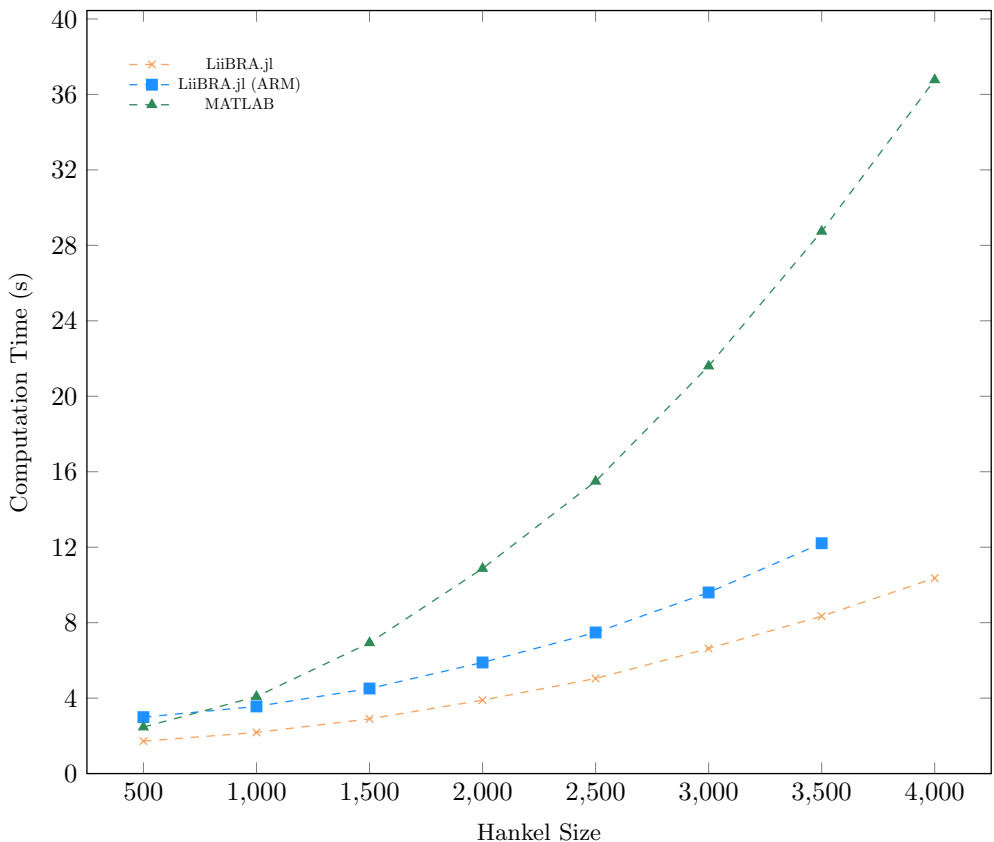
\vspace{1em}

\subsection{\label{sec:Exp_Results} Experimental Validation}
{Experimental validation of LiiBRA.jl's capabilities is presented through parameterisation of an LG Chem. M50 cylindrical 21700 cell\cite{chen_2020}. This is seen as a widely available cell that provides a strong reference for the current state of a high energy conventional intercalation cell with an NCM 811 positive electrode and bi-component $\text{Graphite-SiO}_x$ negative electrode. To the authors knowledge, the discrete realisation algorithm (DRA) has been validated from full-order and linearised partial differential implementations\cite{lee_one-dimensional_2012, rodriguez_comparing_2019, rodriguez_improved_2018, gopalakrishnan_fast_2017}; however, an experimental validation has not been presented in literature. In this section, an experimental voltage validation of the CI-DRA utilising LiiBRA.jl and the parameterised LG Chem. M50 dataset is presented.\\} 

{For this validation, three cells were experimentally tested to reduce cell-to-cell variance. This is seen as the minimal requirement and future investigations are recommended to verify the minimum number of experimentally tested cells required to capture adequate statical variations\cite{dechent_estimation_2021} with respect to LiiBRA.jl . Each cell is initially conditioned at 25\degree C for five cycles at a 1C discharge rate and a $C/2$ charge rate utilising an Arbin LBT21084 cycler and a Binder KB115 incubator. This is followed by a discharge to 75\% SOC and a WLTP drive-cycle is performed based on the specifications shown in Table \ref{WLTP Drive-cycle}. A T-type thermocouple is surface mounted with thermal paste at body centre of the cell to ensure data consistency. \\}

{This data was then compared to both LiiBRA.jl and a Doyle-Fuller-Newman implementation utilising PyBaMM \cite{Sulzer_Python_Battery_Mathematical_2021}. Figure \ref{WLTP_Voltage_Comp} below showcases the predicted voltages for both models and an experimentally measured cell for the WLTP drive-cycle. These results verify the capabilities of the CI-DRA algorithm and LiiBRA.jl for physics-based predictions with prediction values producing a root mean square deviation to PyBaMM of 3.67mV and 7.54mV to the experimental cell. It should be noted that both PyBaMM and LiiBRA.jl experience increased cell voltage error throughout the length of the drive-cycle. This is believed to result from the variation between the experimental and modelled applied current, and additional error in the experimental channel calibration. Online SOC estimation for the conventional DRA has already been presented,\cite{stetzel_electrochemical_2015} which provides a viable correction for this longer-term deviation. \\}

\vspace{1em}
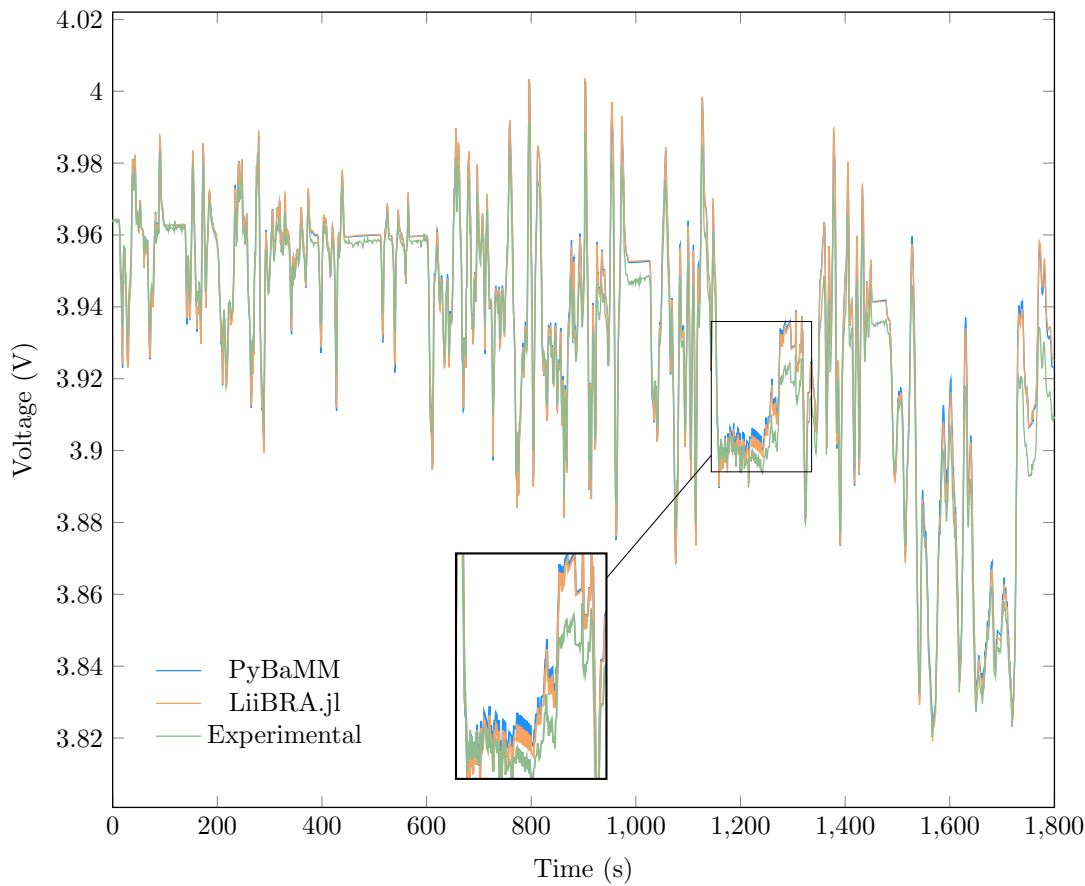
\begin{figure}[H]
\centering
\begin{tikzpicture}[spy using outlines=
	{rectangle, magnification=1.5, connect spies}]
	
    \begin{axis}[
  	width=0.85\linewidth,
  	line width=0.5,
  	grid=major, 
  	tick label style={font=\normalsize},
	xmin=0,xmax=1800,
	grid style={white},
	ylabel style = {align=center},
        ylabel={Voltage (V)},
        xlabel=Time (s),
        tick label style={font=\normalsize},
        legend style={at={(0.28,0.2)},anchor=north east,  draw=none, fill=none},
        each nth point=4, filter discard warning=false, unbounded coords=discard]
        
        		\addplot [color=DodgerBlue] table [x=PyBaMM_T, y=PyBaMM_V, col sep=comma] {WLTP.csv};
		\addplot [color=SandyBrown]table[x=LiiBRA_T, y=LiiBRA_V, col sep=comma] {WLTP.csv};
        		\addplot [color=DarkSeaGreen] table [x=Experimental_T, y=Experimental_V, col sep=comma] {WLTP.csv};
		
		\addlegendimage{/pgfplots/refstyle=plot2}\addlegendentry{PyBaMM}
    		\addlegendimage{/pgfplots/refstyle=plot1}\addlegendentry{LiiBRA.jl}
    		\addlegendimage{/pgfplots/refstyle=plot3}\addlegendentry{Experimental}
	
	 \coordinate (spypoint) at (axis cs:1240,3.915);
  	 \coordinate (magnifyglass) at (axis cs:800,3.84);
\end{axis}

\spy [black, width=2cm,height=3cm] on (spypoint)
   in node[fill=white] at (magnifyglass);
   
\end{tikzpicture}
\caption{Voltage Comparison of Experimental Cell, LiiBRA.jl, and PyBAMM DFN at 75\% SOC WLTP}
\label{WLTP_Voltage_Comp}
\end{figure}
\vspace{1em}

\section{\label{sec:Conclusion}Conclusion}

{This paper presents an open-source modelling package, LiiBRA.jl, developed in Julia for creation and simulation of real-time capable electrochemical models. An expanded realisation algorithm (CI-DRA) is presented with computational implementation discussed and resulting showing improvements over the conventional method. This paper shows capabilities in both conventional offline creation as well as expansion into in-vehicle creation via ARM compilation. This advancement was feasible due to the high-performance capabilities of LiiBRA.jl, and ease of ARM based compilation offered by the Julia language. This improvement opens future parameterisation of in-vehicle online models, providing a vital mechanism for individualised pack degradation over the vehicle's life. This package provides a mean value improvement over the conventional Matlab DRA implementation of 88\%. For ARM deployment, this package provides a viable solution for in-vehicle model generation, with a modest 43\% decrease in performance when compared to an equivalent x86 characterisation. Investigations showed a computational solution time of 5.5 seconds per model for ARM based generation, when LiiBRA.jl was parameterised equivalently to the results shown in Figure \ref{Cs-Neg-Comparision}.\\ 

An investigation into the discrete realisation algorithms parameterisation was completed, with a sensitivity analysis presented. This was continued into experimental validation of the LiiBRA.jl implemented model, with voltage prediction of a WLTP drive cycle resulting in a RMSE value of 3.67mV. Lastly, a methodology for selected the input parameterisation was presented with a solution shown for multiple applications. Utilising Julia for this work has provided a performant solution to the two language problem in real-time embedded computing further reducing resources in software creation and maintenance.\\

Further work in aligning this Julia package for easy utilisation is planned, in addition to further improvements in the realisation algorithms. Furthermore, implementation of a memory efficient singular value decomposition similar to the method presented in Gopalakrishnan et al.\cite{gopalakrishnan_fast_2017} is planned. Lastly, degradation coupling for in-vehicle degradation informed predictions are currently being investigated.\\}

\section*{Appendix A}

{To create the worldwide harmonised light vehicle test procedure (WLTP) for validation of LiiBRA.jl, a model capturing the longitude vehicle dynamics was created. This model utilised parameter specifications from a Tesla Model 3 long range vehicle. These parameters are shown in Table \ref{WLTP Drive-cycle} below. It should be noted that this parameterisation is intended as a reference, and not to exactly match the presented vehicle. Additionally, $^*$ denotes a fitting variable to achieve the reported WLTP range and stored pack energy.\\}
\setcounter{table}{0}
\renewcommand{\thetable}{A\arabic{table}}

\vspace{1em}
\begin{table}[H]
\caption{Tesla Model 3 Long Range specifications used for simulated WLTP}
\centering
\label{WLTP Drive-cycle}
\begin{tabular}{ccccc}
Variable & Definition & Value & Unit\\ \hline
$M$ & Total Vehicle Mass & 1931 & kg\\
$E$ &  Onboard Useable Energy & 82 & kWh \\
$N_s / N_p$ & Electric System Orientation & 96s47p$^*$ & -\\
$C_{Nom}$ & Rated Single Cell Capacity & 5 & Ahr \\
$V_{Lim}$& Operational Voltage Limits& 2.5 / 4.2 & V \\
$F_{D}$& Vehicle Drivetrain Losses  & \cite{Epa-Losses} & N \\
$\eta_M$ & Motor Efficiency  & 0.827$^*$ & - \\
\end{tabular}
\end{table}
\vspace{2em}

\bibliography{LiiBRA}

\providecommand{\noopsort}[1]{}\providecommand{\singleletter}[1]{#1}%
\begin{thebibliography}{10}
\expandafter\ifx\csname url\endcsname\relax
  \def\url#1{\texttt{#1}}\fi
\expandafter\ifx\csname urlprefix\endcsname\relax\def\urlprefix{URL }\fi
\expandafter\ifx\csname href\endcsname\relax
  \def\href#1#2{#2} \def\path#1{#1}\fi

\bibitem{edge_lithium_2021}
J.~S. Edge, S.~O’Kane, R.~Prosser, N.~D. Kirkaldy, A.~N. Patel, A.~Hales,
  A.~Ghosh, W.~Ai, J.~Chen, J.~Yang, S.~Li, M.-C. Pang, L.~Bravo~Diaz,
  A.~Tomaszewska, M.~W. Marzook, K.~N. Radhakrishnan, H.~Wang, Y.~Patel, B.~Wu,
  G.~J. Offer, \href{http://xlink.rsc.org/?DOI=D1CP00359C}{Lithium ion battery
  degradation: what you need to know}, Physical Chemistry Chemical Physics
  23~(14) (2021) 8200--8221.
\newblock \href {http://dx.doi.org/10.1039/D1CP00359C}
  {\path{doi:10.1039/D1CP00359C}}.
\newline\urlprefix\url{http://xlink.rsc.org/?DOI=D1CP00359C}

\bibitem{zhao_observability_2017}
S.~Zhao, S.~R. Duncan, D.~A. Howey,
  \href{http://ieeexplore.ieee.org/document/7464285/}{Observability {Analysis}
  and {State} {Estimation} of {Lithium}-{Ion} {Batteries} in the {Presence} of
  {Sensor} {Biases}}, IEEE Transactions on Control Systems Technology 25~(1)
  (2017) 326--333.
\newblock \href {http://dx.doi.org/10.1109/TCST.2016.2542115}
  {\path{doi:10.1109/TCST.2016.2542115}}.
\newline\urlprefix\url{http://ieeexplore.ieee.org/document/7464285/}

\bibitem{hu_comparative_2012}
X.~Hu, S.~Li, H.~Peng,
  \href{https://linkinghub.elsevier.com/retrieve/pii/S0378775311019628}{A
  comparative study of equivalent circuit models for {Li}-ion batteries},
  Journal of Power Sources 198 (2012) 359--367.
\newblock \href {http://dx.doi.org/10.1016/j.jpowsour.2011.10.013}
  {\path{doi:10.1016/j.jpowsour.2011.10.013}}.
\newline\urlprefix\url{https://linkinghub.elsevier.com/retrieve/pii/S0378775311019628}

\bibitem{plett_battery_2015}
G.~L. Plett, Battery management systems: battery modeling. {Volume} 1, Artech
  House, Boston : London, 2015, oCLC: ocn909081842.

\bibitem{petit_development_2016}
M.~Petit, E.~Prada, V.~Sauvant-Moynot,
  \href{https://linkinghub.elsevier.com/retrieve/pii/S0306261916304500}{Development
  of an empirical aging model for {Li}-ion batteries and application to assess
  the impact of {Vehicle}-to-{Grid} strategies on battery lifetime}, Applied
  Energy 172 (2016) 398--407.
\newblock \href {http://dx.doi.org/10.1016/j.apenergy.2016.03.119}
  {\path{doi:10.1016/j.apenergy.2016.03.119}}.
\newline\urlprefix\url{https://linkinghub.elsevier.com/retrieve/pii/S0306261916304500}

\bibitem{de_hoog_combined_2017}
J.~de~Hoog, J.-M. Timmermans, D.~Ioan-Stroe, M.~Swierczynski, J.~Jaguemont,
  S.~Goutam, N.~Omar, J.~Van~Mierlo, P.~Van Den~Bossche,
  \href{https://linkinghub.elsevier.com/retrieve/pii/S0306261917305251}{Combined
  cycling and calendar capacity fade modeling of a
  {Nickel}-{Manganese}-{Cobalt} {Oxide} {Cell} with real-life profile
  validation}, Applied Energy 200 (2017) 47--61.
\newblock \href {http://dx.doi.org/10.1016/j.apenergy.2017.05.018}
  {\path{doi:10.1016/j.apenergy.2017.05.018}}.
\newline\urlprefix\url{https://linkinghub.elsevier.com/retrieve/pii/S0306261917305251}

\bibitem{reniers_improving_2018}
J.~M. Reniers, G.~Mulder, S.~Ober-Blöbaum, D.~A. Howey,
  \href{https://linkinghub.elsevier.com/retrieve/pii/S0378775318300041}{Improving
  optimal control of grid-connected lithium-ion batteries through more accurate
  battery and degradation modelling}, Journal of Power Sources 379 (2018)
  91--102.
\newblock \href {http://dx.doi.org/10.1016/j.jpowsour.2018.01.004}
  {\path{doi:10.1016/j.jpowsour.2018.01.004}}.
\newline\urlprefix\url{https://linkinghub.elsevier.com/retrieve/pii/S0378775318300041}

\bibitem{schimpe_comprehensive_2018}
M.~Schimpe, M.~E. von Kuepach, M.~Naumann, H.~C. Hesse, K.~Smith, A.~Jossen,
  \href{https://iopscience.iop.org/article/10.1149/2.1181714jes}{Comprehensive
  {Modeling} of {Temperature}-{Dependent} {Degradation} {Mechanisms} in
  {Lithium} {Iron} {Phosphate} {Batteries}}, Journal of The Electrochemical
  Society 165~(2) (2018) A181--A193.
\newblock \href {http://dx.doi.org/10.1149/2.1181714jes}
  {\path{doi:10.1149/2.1181714jes}}.
\newline\urlprefix\url{https://iopscience.iop.org/article/10.1149/2.1181714jes}

\bibitem{doyle_modeling_1993}
M.~Doyle, \href{https://iopscience.iop.org/article/10.1149/1.2221597}{Modeling
  of {Galvanostatic} {Charge} and {Discharge} of the
  {Lithium}/{Polymer}/{Insertion} {Cell}}, Journal of The Electrochemical
  Society 140~(6) (1993) 1526.
\newblock \href {http://dx.doi.org/10.1149/1.2221597}
  {\path{doi:10.1149/1.2221597}}.
\newline\urlprefix\url{https://iopscience.iop.org/article/10.1149/1.2221597}

\bibitem{fuller_simulation_1994}
T.~F. Fuller, M.~Doyle, J.~Newman,
  \href{https://iopscience.iop.org/article/10.1149/1.2054684}{Simulation and
  {Optimization} of the {Dual} {Lithium} {Ion} {Insertion} {Cell}}, Journal of
  The Electrochemical Society 141~(1) (1994) 1--10.
\newblock \href {http://dx.doi.org/10.1149/1.2054684}
  {\path{doi:10.1149/1.2054684}}.
\newline\urlprefix\url{https://iopscience.iop.org/article/10.1149/1.2054684}

\bibitem{marquis_asymptotic_2019}
S.~G. Marquis, V.~Sulzer, R.~Timms, C.~P. Please, S.~J. Chapman,
  \href{http://arxiv.org/abs/1905.12553}{An asymptotic derivation of a single
  particle model with electrolyte}, arXiv:1905.12553 [physics]ArXiv:
  1905.12553.
\newline\urlprefix\url{http://arxiv.org/abs/1905.12553}

\bibitem{jang_towards_2021}
T.~Jang, L.~Mishra, K.~Shah, A.~Subramaniam, M.~Uppaluri, S.~A. Roberts, V.~R.
  Subramanian,
  \href{https://iopscience.iop.org/article/10.1149/10401.0131ecst}{Towards
  {Real}-{Time} {Simulation} of {Two}-{Dimensional} {Models} for
  {Electrodeposition}/{Stripping} in {Lithium}-{Metal} {Batteries}}, ECS
  Transactions 104~(1) (2021) 131--152.
\newblock \href {http://dx.doi.org/10.1149/10401.0131ecst}
  {\path{doi:10.1149/10401.0131ecst}}.
\newline\urlprefix\url{https://iopscience.iop.org/article/10.1149/10401.0131ecst}

\bibitem{mishra_perspectivemass_2021}
L.~Mishra, A.~Subramaniam, T.~Jang, K.~Shah, M.~Uppaluri, S.~A. Roberts, V.~R.
  Subramanian,
  \href{https://iopscience.iop.org/article/10.1149/1945-7111/ac2091}{Perspective—{Mass}
  {Conservation} in {Models} for {Electrodeposition}/{Stripping} in {Lithium}
  {Metal} {Batteries}}, Journal of The Electrochemical Society 168~(9) (2021)
  092502.
\newblock \href {http://dx.doi.org/10.1149/1945-7111/ac2091}
  {\path{doi:10.1149/1945-7111/ac2091}}.
\newline\urlprefix\url{https://iopscience.iop.org/article/10.1149/1945-7111/ac2091}

\bibitem{reniers_review_2019}
J.~M. Reniers, G.~Mulder, D.~A. Howey,
  \href{https://iopscience.iop.org/article/10.1149/2.0281914jes}{Review and
  {Performance} {Comparison} of {Mechanical}-{Chemical} {Degradation} {Models}
  for {Lithium}-{Ion} {Batteries}}, Journal of The Electrochemical Society
  166~(14) (2019) A3189--A3200, number: 14.
\newblock \href {http://dx.doi.org/10.1149/2.0281914jes}
  {\path{doi:10.1149/2.0281914jes}}.
\newline\urlprefix\url{https://iopscience.iop.org/article/10.1149/2.0281914jes}

\bibitem{Forman_2011}
J.~C. Forman, S.~Bashash, J.~L. Stein, H.~K. Fathy,
  \href{https://doi.org/10.1149/1.3519059}{Reduction of an
  electrochemistry-based li-ion battery model via quasi-linearization and
  pad\'e approximation}, Journal of The Electrochemical Society 158~(2) (2011)
  A93.
\newblock \href {http://dx.doi.org/10.1149/1.3519059}
  {\path{doi:10.1149/1.3519059}}.
\newline\urlprefix\url{https://doi.org/10.1149/1.3519059}

\bibitem{smith_model_2008}
K.~A. Smith, C.~D. Rahn, C.-Y. Wang,
  \href{https://doi.org/10.1115/1.2807068}{Model {Order} {Reduction} of {1D}
  {Diffusion} {Systems} {Via} {Residue} {Grouping}}, Journal of Dynamic
  Systems, Measurement, and Control 130~(1).
\newblock \href {http://dx.doi.org/10.1115/1.2807068}
  {\path{doi:10.1115/1.2807068}}.
\newline\urlprefix\url{https://doi.org/10.1115/1.2807068}

\bibitem{Ramadesigan_2010}
V.~Ramadesigan, V.~Boovaragavan, J.~C. Pirkle, V.~R. Subramanian,
  \href{https://doi.org/10.1149/1.3425622}{Efficient reformulation of
  solid-phase diffusion in physics-based lithium-ion battery models}, Journal
  of The Electrochemical Society 157~(7) (2010) A854.
\newblock \href {http://dx.doi.org/10.1149/1.3425622}
  {\path{doi:10.1149/1.3425622}}.
\newline\urlprefix\url{https://doi.org/10.1149/1.3425622}

\bibitem{Subramanian_2005}
V.~R. Subramanian, V.~D. Diwakar, D.~Tapriyal,
  \href{https://doi.org/10.1149/1.2032427}{Efficient macro-micro scale coupled
  modeling of batteries}, Journal of The Electrochemical Society 152~(10)
  (2005) A2002.
\newblock \href {http://dx.doi.org/10.1149/1.2032427}
  {\path{doi:10.1149/1.2032427}}.
\newline\urlprefix\url{https://doi.org/10.1149/1.2032427}

\bibitem{lee_discrete-time_2012}
J.~L. Lee, A.~Chemistruck, G.~L. Plett,
  \href{https://linkinghub.elsevier.com/retrieve/pii/S0378775312002789}{Discrete-time
  realization of transcendental impedance models, with application to modeling
  spherical solid diffusion}, Journal of Power Sources 206 (2012) 367--377.
\newblock \href {http://dx.doi.org/10.1016/j.jpowsour.2012.01.134}
  {\path{doi:10.1016/j.jpowsour.2012.01.134}}.
\newline\urlprefix\url{https://linkinghub.elsevier.com/retrieve/pii/S0378775312002789}

\bibitem{rodriguez_comparing_2019}
A.~Rodríguez, G.~L. Plett, M.~S. Trimboli,
  \href{https://linkinghub.elsevier.com/retrieve/pii/S2590116819300098}{Comparing
  four model-order reduction techniques, applied to lithium-ion battery-cell
  internal electrochemical transfer functions}, eTransportation 1 (2019)
  100009.
\newblock \href {http://dx.doi.org/10.1016/j.etran.2019.100009}
  {\path{doi:10.1016/j.etran.2019.100009}}.
\newline\urlprefix\url{https://linkinghub.elsevier.com/retrieve/pii/S2590116819300098}

\bibitem{jin_physically-based_2017}
X.~Jin, A.~Vora, V.~Hoshing, T.~Saha, G.~Shaver, R.~E. García, O.~Wasynczuk,
  S.~Varigonda,
  \href{https://linkinghub.elsevier.com/retrieve/pii/S037877531631802X}{Physically-based
  reduced-order capacity loss model for graphite anodes in {Li}-ion battery
  cells}, Journal of Power Sources 342 (2017) 750--761.
\newblock \href {http://dx.doi.org/10.1016/j.jpowsour.2016.12.099}
  {\path{doi:10.1016/j.jpowsour.2016.12.099}}.
\newline\urlprefix\url{https://linkinghub.elsevier.com/retrieve/pii/S037877531631802X}

\bibitem{2021JPS...49029571H}
S.~{Han}, Y.~{Tang}, S.~{Khaleghi Rahimian}, {A numerically efficient method of
  solving the full-order pseudo-2-dimensional (P2D) Li-ion cell model}, Journal
  of Power Sources 490 (2021) 229571.
\newblock \href {http://dx.doi.org/10.1016/j.jpowsour.2021.229571}
  {\path{doi:10.1016/j.jpowsour.2021.229571}}.

\bibitem{SMITH2006662}
K.~Smith, C.-Y. Wang,
  \href{https://www.sciencedirect.com/science/article/pii/S0378775306001017}{Power
  and thermal characterization of a lithium-ion battery pack for
  hybrid-electric vehicles}, Journal of Power Sources 160~(1) (2006) 662--673.
\newblock \href
  {http://dx.doi.org/https://doi.org/10.1016/j.jpowsour.2006.01.038}
  {\path{doi:https://doi.org/10.1016/j.jpowsour.2006.01.038}}.
\newline\urlprefix\url{https://www.sciencedirect.com/science/article/pii/S0378775306001017}

\bibitem{Comp_Framework_MPC}
M.~A. Xavier, A.~K. de~Souza, K.~Karami, G.~L. Plett, M.~S. Trimboli, A
  computational framework for lithium ion cell-level model predictive control
  using a physics-based reduced-order model, IEEE Control Systems Letters 5~(4)
  (2021) 1387--1392.
\newblock \href {http://dx.doi.org/10.1109/LCSYS.2020.3038131}
  {\path{doi:10.1109/LCSYS.2020.3038131}}.

\bibitem{Julia-2017}
J.~Bezanson, A.~Edelman, S.~Karpinski, V.~B. Shah,
  \href{https://epubs.siam.org/doi/10.1137/141000671}{Julia: A fresh approach
  to numerical computing}, SIAM Review 59~(1) (2017) 65--98.
\newblock \href {http://dx.doi.org/10.1137/141000671}
  {\path{doi:10.1137/141000671}}.
\newline\urlprefix\url{https://epubs.siam.org/doi/10.1137/141000671}

\bibitem{newman_porous-electrode_1975}
J.~Newman, W.~Tiedemann,
  \href{https://onlinelibrary.wiley.com/doi/10.1002/aic.690210103}{Porous-electrode
  theory with battery applications}, AIChE Journal 21~(1) (1975) 25--41.
\newblock \href {http://dx.doi.org/10.1002/aic.690210103}
  {\path{doi:10.1002/aic.690210103}}.
\newline\urlprefix\url{https://onlinelibrary.wiley.com/doi/10.1002/aic.690210103}

\bibitem{newman2012electrochemical}
J.~Newman, K.~E. Thomas-Alyea, Electrochemical systems, John Wiley \& Sons,
  2012.

\bibitem{smith_solid-state_2006}
K.~Smith, C.-Y. Wang,
  \href{https://linkinghub.elsevier.com/retrieve/pii/S0378775306006161}{Solid-state
  diffusion limitations on pulse operation of a lithium ion cell for hybrid
  electric vehicles}, Journal of Power Sources 161~(1) (2006) 628--639.
\newblock \href {http://dx.doi.org/10.1016/j.jpowsour.2006.03.050}
  {\path{doi:10.1016/j.jpowsour.2006.03.050}}.
\newline\urlprefix\url{https://linkinghub.elsevier.com/retrieve/pii/S0378775306006161}

\bibitem{subramanian_mathematical_2009}
V.~R. Subramanian, V.~Boovaragavan, V.~Ramadesigan, M.~Arabandi,
  \href{https://iopscience.iop.org/article/10.1149/1.3065083}{Mathematical
  {Model} {Reformulation} for {Lithium}-{Ion} {Battery} {Simulations}:
  {Galvanostatic} {Boundary} {Conditions}}, Journal of The Electrochemical
  Society 156~(4) (2009) A260.
\newblock \href {http://dx.doi.org/10.1149/1.3065083}
  {\path{doi:10.1149/1.3065083}}.
\newline\urlprefix\url{https://iopscience.iop.org/article/10.1149/1.3065083}

\bibitem{bizeray_lithium-ion_2015}
A.~M. Bizeray, S.~Zhao, S.~R. Duncan, D.~A. Howey,
  \href{http://arxiv.org/abs/1506.08689}{Lithium-ion battery
  thermal-electrochemical model-based state estimation using orthogonal
  collocation and a modified extended {Kalman} filter}, Journal of Power
  Sources 296 (2015) 400--412, arXiv: 1506.08689.
\newblock \href {http://dx.doi.org/10.1016/j.jpowsour.2015.07.019}
  {\path{doi:10.1016/j.jpowsour.2015.07.019}}.
\newline\urlprefix\url{http://arxiv.org/abs/1506.08689}

\bibitem{cai_lithium_2012}
L.~Cai, R.~E. White,
  \href{https://linkinghub.elsevier.com/retrieve/pii/S0378775312010439}{Lithium
  ion cell modeling using orthogonal collocation on finite elements}, Journal
  of Power Sources 217 (2012) 248--255.
\newblock \href {http://dx.doi.org/10.1016/j.jpowsour.2012.06.043}
  {\path{doi:10.1016/j.jpowsour.2012.06.043}}.
\newline\urlprefix\url{https://linkinghub.elsevier.com/retrieve/pii/S0378775312010439}

\bibitem{northrop_coordinate_2011}
P.~W.~C. Northrop, V.~Ramadesigan, S.~De, V.~R. Subramanian,
  \href{https://iopscience.iop.org/article/10.1149/2.058112jes}{Coordinate
  {Transformation}, {Orthogonal} {Collocation}, {Model} {Reformulation} and
  {Simulation} of {Electrochemical}-{Thermal} {Behavior} of {Lithium}-{Ion}
  {Battery} {Stacks}}, Journal of The Electrochemical Society 158~(12) (2011)
  A1461.
\newblock \href {http://dx.doi.org/10.1149/2.058112jes}
  {\path{doi:10.1149/2.058112jes}}.
\newline\urlprefix\url{https://iopscience.iop.org/article/10.1149/2.058112jes}

\bibitem{drummond_feedback_2020}
R.~Drummond, A.~M. Bizeray, D.~A. Howey, S.~R. Duncan,
  \href{https://ieeexplore.ieee.org/document/8694827/}{A {Feedback}
  {Interpretation} of the {Doyle}–{Fuller}–{Newman} {Lithium}-{Ion}
  {Battery} {Model}}, IEEE Transactions on Control Systems Technology 28~(4)
  (2020) 1284--1295.
\newblock \href {http://dx.doi.org/10.1109/TCST.2019.2909722}
  {\path{doi:10.1109/TCST.2019.2909722}}.
\newline\urlprefix\url{https://ieeexplore.ieee.org/document/8694827/}

\bibitem{jacobsen_diffusion_1995}
T.~Jacobsen, K.~West,
  \href{https://linkinghub.elsevier.com/retrieve/pii/0013468694E01923}{Diffusion
  impedance in planar, cylindrical and spherical symmetry}, Electrochimica Acta
  40~(2) (1995) 255--262.
\newblock \href {http://dx.doi.org/10.1016/0013-4686(94)E0192-3}
  {\path{doi:10.1016/0013-4686(94)E0192-3}}.
\newline\urlprefix\url{https://linkinghub.elsevier.com/retrieve/pii/0013468694E01923}

\bibitem{smith_control_2007}
K.~A. Smith, C.~D. Rahn, C.-Y. Wang,
  \href{https://linkinghub.elsevier.com/retrieve/pii/S0196890407000908}{Control
  oriented {1D} electrochemical model of lithium ion battery}, Energy
  Conversion and Management 48~(9) (2007) 2565--2578.
\newblock \href {http://dx.doi.org/10.1016/j.enconman.2007.03.015}
  {\path{doi:10.1016/j.enconman.2007.03.015}}.
\newline\urlprefix\url{https://linkinghub.elsevier.com/retrieve/pii/S0196890407000908}

\bibitem{lee_one-dimensional_2012}
J.~L. Lee, A.~Chemistruck, G.~L. Plett,
  \href{https://linkinghub.elsevier.com/retrieve/pii/S0378775312012104}{One-dimensional
  physics-based reduced-order model of lithium-ion dynamics}, Journal of Power
  Sources 220 (2012) 430--448.
\newblock \href {http://dx.doi.org/10.1016/j.jpowsour.2012.07.075}
  {\path{doi:10.1016/j.jpowsour.2012.07.075}}.
\newline\urlprefix\url{https://linkinghub.elsevier.com/retrieve/pii/S0378775312012104}

\bibitem{oppenheim2001discrete}
A.~V. Oppenheim, J.~R. Buck, R.~W. Schafer, Discrete-time signal processing.
  Vol. 2, Upper Saddle River, NJ: Prentice Hall, 2001.

\bibitem{ho_effective_1966}
B.~HO, R.~E. K{\'a}lm{\'a}n, Effective construction of linear state-variable
  models from input/output functions, at-Automatisierungstechnik 14~(1-12)
  (1966) 545--548.

\bibitem{larsen_lanczos_1998}
R.~M. Larsen, \href{https://tidsskrift.dk/daimipb/article/view/7070}{Lanczos
  {Bidiagonalization} {With} {Partial} {Reorthogonalization}}, DAIMI Report
  Series 27~(537).
\newblock \href {http://dx.doi.org/10.7146/dpb.v27i537.7070}
  {\path{doi:10.7146/dpb.v27i537.7070}}.
\newline\urlprefix\url{https://tidsskrift.dk/daimipb/article/view/7070}

\bibitem{dierckx1995curve}
P.~Dierckx, \href{https://github.com/kbarbary/Dierckx.jl}{Curve and surface
  fitting with splines}, Oxford University Press, 1995.
\newline\urlprefix\url{https://github.com/kbarbary/Dierckx.jl}

\bibitem{FFTW.jl-2005}
M.~Frigo, S.~G. Johnson, The design and implementation of {FFTW3}, Proceedings
  of the IEEE 93~(2) (2005) 216--231, special issue on ``Program Generation,
  Optimization, and Platform Adaptation''.
\newblock \href {http://dx.doi.org/10.1109/JPROC.2004.840301}
  {\path{doi:10.1109/JPROC.2004.840301}}.

\bibitem{Lehoucq97arpackusers}
R.~B. Lehoucq, D.~C. Sorensen, C.~Yang, Arpack users guide: Solution of large
  scale eigenvalue problems by implicitly restarted arnoldi methods. (1997).

\bibitem{dominique_2021_5090075}
Dominique, Alexis, A.~Noack, JSOBot, J.~Chen, MonssafToukal, A.~Siqueira,
  J.~TagBot,
  \href{https://doi.org/10.5281/zenodo.5090075}{Juliasmoothoptimizers/propack.jl:
  v0.4.0} (Jul. 2021).
\newblock \href {http://dx.doi.org/10.5281/zenodo.5090075}
  {\path{doi:10.5281/zenodo.5090075}}.
\newline\urlprefix\url{https://doi.org/10.5281/zenodo.5090075}

\bibitem{arnoldi1951principle}
W.~E. Arnoldi, The principle of minimized iterations in the solution of the
  matrix eigenvalue problem, Quarterly of applied mathematics 9~(1) (1951)
  17--29.

\bibitem{BenchmarkTools.jl-2016}
J.~{Chen}, J.~{Revels}, {Robust benchmarking in noisy environments}, arXiv
  e-prints\href {http://arxiv.org/abs/1608.04295} {\path{arXiv:1608.04295}}.

\bibitem{WLTP_2021}
{United Nations Economic Commission for Europe}, Global technical regulation -
  ece/trans/180/add.15/amend.6,
  \newline\url{https://unece.org/transport/standards/transport/vehicle-regulations-wp29/global-technical-regulations-gtrs}
  (2021).

\bibitem{Sulzer_Python_Battery_Mathematical_2021}
V.~Sulzer, S.~G. Marquis, R.~Timms, M.~Robinson, S.~J. Chapman,
  \href{https://github.com/pybamm-team/PyBaMM}{Python battery mathematical
  modelling (pybamm)} (6 2021).
\newblock \href {http://dx.doi.org/10.5334/jors.309}
  {\path{doi:10.5334/jors.309}}.
\newline\urlprefix\url{https://github.com/pybamm-team/PyBaMM}

\bibitem{chen_2020}
C.-H. Chen, F.~Brosa~Planella, K.~O’Regan, D.~Gastol, W.~D. Widanage,
  E.~Kendrick,
  \href{https://iopscience.iop.org/article/10.1149/1945-7111/ab9050}{Development
  of experimental techniques for parameterization of multi-scale lithium-ion
  battery models}, Journal of The Electrochemical Society 167~(8) (2020)
  080534.
\newblock \href {http://dx.doi.org/10.1149/1945-7111/ab9050}
  {\path{doi:10.1149/1945-7111/ab9050}}.
\newline\urlprefix\url{https://iopscience.iop.org/article/10.1149/1945-7111/ab9050}

\bibitem{rodriguez_improved_2018}
A.~Rodríguez, G.~L. Plett, M.~S. Trimboli,
  \href{https://linkinghub.elsevier.com/retrieve/pii/S2352152X18302032}{Improved
  transfer functions modeling linearized lithium-ion battery-cell internal
  electrochemical variables}, Journal of Energy Storage 20 (2018) 560--575.
\newblock \href {http://dx.doi.org/10.1016/j.est.2018.06.015}
  {\path{doi:10.1016/j.est.2018.06.015}}.
\newline\urlprefix\url{https://linkinghub.elsevier.com/retrieve/pii/S2352152X18302032}

\bibitem{gopalakrishnan_fast_2017}
K.~Gopalakrishnan, T.~Zhang, G.~J. Offer,
  \href{https://asmedigitalcollection.asme.org/electrochemical/article/doi/10.1115/1.4035526/380537/A-Fast-MemoryEfficient-DiscreteTime-Realization}{A
  {Fast}, {Memory}-{Efficient} {Discrete}-{Time} {Realization} {Algorithm} for
  {Reduced}-{Order} {Li}-{Ion} {Battery} {Models}}, Journal of Electrochemical
  Energy Conversion and Storage 14~(1) (2017) 011001.
\newblock \href {http://dx.doi.org/10.1115/1.4035526}
  {\path{doi:10.1115/1.4035526}}.
\newline\urlprefix\url{https://asmedigitalcollection.asme.org/electrochemical/article/doi/10.1115/1.4035526/380537/A-Fast-MemoryEfficient-DiscreteTime-Realization}

\bibitem{dechent_estimation_2021}
P.~Dechent, S.~Greenbank, F.~Hildenbrand, S.~Jbabdi, D.~U. Sauer, D.~A. Howey,
  \href{https://onlinelibrary.wiley.com/doi/10.1002/batt.202100148}{Estimation
  of {Li}‐{Ion} {Degradation} {Test} {Sample} {Sizes} {Required} to
  {Understand} {Cell}‐to‐{Cell} {Variability}**}, Batteries \& Supercaps
  4~(12) (2021) 1821--1829.
\newblock \href {http://dx.doi.org/10.1002/batt.202100148}
  {\path{doi:10.1002/batt.202100148}}.
\newline\urlprefix\url{https://onlinelibrary.wiley.com/doi/10.1002/batt.202100148}

\bibitem{stetzel_electrochemical_2015}
K.~D. Stetzel, L.~L. Aldrich, M.~S. Trimboli, G.~L. Plett,
  \href{https://linkinghub.elsevier.com/retrieve/pii/S0378775314020023}{Electrochemical
  state and internal variables estimation using a reduced-order physics-based
  model of a lithium-ion cell and an extended {Kalman} filter}, Journal of
  Power Sources 278 (2015) 490--505.
\newblock \href {http://dx.doi.org/10.1016/j.jpowsour.2014.11.135}
  {\path{doi:10.1016/j.jpowsour.2014.11.135}}.
\newline\urlprefix\url{https://linkinghub.elsevier.com/retrieve/pii/S0378775314020023}

\bibitem{Epa-Losses}
{U.S. Environmental Protection Agency}, 2022 tesla model 3 long range awd -
  certification summary information report (2022).

\end{thebibliography}

\end{document}